\documentclass[11pt]{article}
\usepackage{cite}
\usepackage{amsmath,amsfonts,amssymb}
\usepackage[small,bf,hang]{caption}
\usepackage{slashed}
\usepackage{mathabx}
\usepackage{latexsym,epsfig}
\usepackage{color}

\def\hybrid{
        \topmargin -20pt
        \oddsidemargin 0pt
        \headheight 0pt \headsep 0pt
        \textwidth 6.25in 
        \textheight 9.5in 
        \marginparwidth .875in
        \parskip 5pt plus 1pt \jot = 1.5ex}

\hybrid

 \csname
@addtoreset\endcsname{equation}{section}

\def\moth{\mathsurround=0pt}
\newdimen\zo \zo=0pt

\def\tick{\leaders\hrule height 0.5ex depth 0pt \hskip 0.5pt}
\def\upboxfill{$\moth \setbox\zo\hbox{\tick}%
  \hskip 3pt\hbox to 0pt{$\tick$\hss}\hrulefill \hbox to 7.5pt{$\tick$\hss}$}

\def\dtick{\leaders\hrule height .34pt depth 0.5ex \hskip 0.5pt}
\def\downboxfill{$\moth \setbox\zo\hbox{\dtick}%
  \hskip 2pt\hbox to 0pt{$\dtick$\hss}\hrulefill \hbox to 2pt{$\dtick$\hss}$}

\def\bec{\begin{center}}
\def\ec{\end{center}}

\def\m{\mu}
\def\n{\nu}

\def\nn{\nonumber}

\def\be{\begin{equation}}
\def\ee{\end{equation}}
\def\bea{\begin{eqnarray}}
\def\eea{\end{eqnarray}}
\def\ba{\begin{array}}
\def\ea{\end{array}}

\def\ft#1#2{{\textstyle{{\scriptstyle #1}
\over {\scriptstyle #2}}}}

\newcommand{\KK}{{_{^{\rm KK}}}}

\begin{document}

\begin{titlepage}
\rightline{}
\rightline{\tt  MIT-CTP/4670}
\rightline{June 2015}
\begin{center}
\vskip .6cm
{\Large \bf { Consistent Type IIB Reductions to Maximal 5D Supergravity}}\\
\vskip 1.2cm
{\large {Arnaud Baguet${}^1$, Olaf Hohm${}^2$ and Henning Samtleben${}^1$}}
\vskip .6cm
{\it {${}^1$Universit\'e de Lyon, Laboratoire de Physique, UMR 5672, CNRS}}\\
{\it {\'Ecole Normale Sup\'erieure de Lyon}}\\
{\it {46, all\'ee d'Italie, F-69364 Lyon cedex 07, France}}\\
arnaud.baguet@ens-lyon.fr, henning.samtleben@ens-lyon.fr
\vskip 0.2cm
{\it {${}^2$Center for Theoretical Physics}}\\
{\it {Massachusetts Institute of Technology}}\\
{\it {Cambridge, MA 02139, USA}}\\
ohohm@mit.edu

\vskip 1cm
{\bf Abstract}
\end{center}

\vskip 0.2cm

\noindent
\begin{narrower}

We use exceptional field theory as a tool to work out the
full non-linear reduction ansaetze for the AdS$_5\times S^5$
compactification of IIB supergravity and its non-compact
counterparts in which the sphere $S^5$ is replaced by the 
inhomogeneous hyperboloidal space $H^{p,q}$.
The resulting theories are the maximal 5D supergravities
with gauge groups ${\rm SO}(p,q)$.
They are consistent truncations in the sense that every solution
of 5D supergravity lifts to a solution of IIB supergravity.
In particular, every stationary point and every holographic RG flow of the scalar potentials
for the compact and non-compact 5D gaugings directly lift
to solutions of IIB supergravity.

\end{narrower}

\vskip 1.5cm

\end{titlepage}

\tableofcontents


\vspace{0.8cm}

\section{Introduction}

It is a notoriously difficult problem to establish the consistency of Kaluza-Klein truncations. 
Consistency requires that any solution of the lower-dimensional theory can be lifted to a solution 
of the original higher-dimensional theory \cite{Duff:1984hn}. 
While this condition is trivially satisfied for torus compactifications, 
the compactification on curved manifolds is generically inconsistent except for very specific
geometries and matter content of the theories. Even in the case of maximally symmetric 
spherical geometries, consistency only holds for a few very special cases~\cite{Cvetic:2000dm}
and even then the proof is often surprisingly laborious. 
An example for a Kaluza-Klein truncation for which a complete proof of consistency 
was out of reach until recently is that of type IIB supergravity on AdS$_5\times {\rm S}^5$, which is  
believed to have a consistent truncation to the maximal SO$(6)$ gauged supergravity in five dimensions 
constructed in \cite{Gunaydin:1984qu,Pernici:1985ju,Gunaydin:1985cu}.
In general not even the form of the non-linear Kaluza-Klein reduction ansatz 
for the higher-dimensional fields is explicitly known, in which case it is 
not even known how to perform the Kaluza-Klein reduction in principle. 
If the reduction ansatz is known 
it remains the task to show that the internal coordinate dependence
of the higher-dimensional field equations factors out such that these equations consistently reduce
to those of the lower-dimensional theory.
Despite these complications, consistency proofs have been obtained over the years for 
various special cases. The maximal eleven-dimensional supergravity admits consistent 
Kaluza-Klein truncations on   AdS$_4\times {\rm S}^7$~\cite{deWit:1986iy}
and AdS$_7\times {\rm S}^4$~\cite{Nastase:1999kf}. 
Subsectors of truncations of type IIB to five dimensions have been 
shown to be consistent in 
\cite{Cvetic:1999xp,Lu:1999bw,Cvetic:2000nc,Pilch:2000ue,Cassani:2010uw,Liu:2010sa,Gauntlett:2010vu,Skenderis:2010vz}. 
More recently, a consistent truncation of massive type IIA supergravity on S$^6$ has been found \cite{Guarino:2015jca}.

In this paper we will present the explicit and complete reduction formulas for a large class of 
truncations of type IIB supergravity to maximal five-dimensional gauged supergravity,
by working out the details of the general construction of \cite{Hohm:2014qga}. 
This includes the famous reduction on AdS$_5\times {\rm S}^5$ to the maximal $D=5$ ${\rm SO}(6)$ gauged
supergravity of \cite{Gunaydin:1985cu}, but also reductions to non-compact gaugings, 
corresponding to truncations with non-compact (hyperboloidal) internal manifolds. Consistency
of the latter has first been conjectured in \cite{Hull:1988jw} and more recently been 
discussed in \cite{Cvetic:2004km,Baron:2014bya}. 
The crucial new ingredient that makes our construction feasible is the recently constructed 
`exceptional field theory' (EFT) \cite{Hohm:2013pua,Hohm:2013vpa,Hohm:2013uia,Hohm:2014fxa} 
and its associated extended geometry, see~\cite{Berman:2010is,Coimbra:2011ky,Aldazabal:2013mya,Hohm:2015xna}, 
and \cite{Siegel:1993th,Hull:2009mi,Hohm:2010jy,Hohm:2010pp} for the closely related double field theory.
Within this framework, the complicated geometric IIB reductions can 
very conveniently be formulated as Scherk-Schwarz reductions on an exceptional 
space-time.

In order to illustrate this point, it is useful to compare it with the toy example 
of an $S^2$ compactification of the $D$-dimensional Einstein-Maxwell theory, whose 
volume form provides the source for the U$(1)$ field strength.
With a particular dilaton coupling, this theory not only permits a vacuum solution with $S^2$ as the compact space 
but also a consistent Kaluza-Klein truncation around this vacuum to a $(D-2)$-dimensional theory~\cite{Cvetic:2000dm}. 
The required dilaton couplings are precisely those that follow from embedding the original theory 
as the $S^1$ reduction of pure gravity in $D+1$ dimensions. 
While the consistency of this reduction can be shown by a direct computation, a far more elegant proof
relies on this geometric origin. As shown in \cite{Cvetic:2003jy}, from the point of view of $(D+1)$ dimensional Einstein 
gravity, the original $S^2$ reduction takes the form of a Scherk-Schwarz (or DeWitt) reduction on a three-dimensional ${\rm SO}(3)$ group
manifold via the Hopf fibration $S^1\hookrightarrow S^3 \rightarrow S^2$\,. 
For Scherk-Schwarz reductions, however, consistency is guaranteed from symmetry arguments~\cite{Scherk:1979zr}, 
which then implies the consistency of the $S^2$ reduction of the Einstein-Maxwell theory.
In this sense, the consistency of the $S^2$ reduction hinges on the fact that the original theory 
is secretly a `geometric' theory in higher dimensions (namely pure Einstein gravity).

Similarly, in exceptional field theory maximal supergravity is reformulated on an 
extended higher-dimensional space that renders the theory covariant w.r.t.~the exceptional 
U-duality groups in the series E$_{d(d)}$, $2\leq d\leq 8$. 
In this case, the higher-dimensional theory is not simply Einstein gravity, but
EFT is subject to a covariant constraint that implies that only a subspace of the extended
space is physical. Solving the constraint accordingly one obtains either type IIB or
eleven-dimensional supergravity. Importantly, the gauge symmetries of EFT are governed 
by `generalized Lie derivatives' that unify the usual diffeomorphism and tensor gauge 
transformations of supergravity into generalized diffeomorphisms of the extended space. 
Specifically, for the E$_{6(6)}$ EFT that will be employed in this paper the generalized Lie derivative 
for vector fields $V^M$, $W^M$, $M,N=1,\ldots, 27$, in the fundamental 
representation $\bar{{\bf 27}}$ reads~\cite{Coimbra:2011ky,Berman:2012vc}
 \be\label{LieIntro}
  \big(\mathbb{L}_{V}W\big)^M \ \equiv \ 
  V^N\partial_NW^M-W^N\partial_N V^M+10\,d^{MNP}\,d_{KLP}\,\partial_NV^K\,W^L\;, 
 \ee  
where $d^{MNK}$ is a (symmetric) invariant tensor of E$_{6(6)}$. Here the first 
two terms represent the standard Lie bracket or derivative on the extended 27-dimensional space, 
while the new term encodes the non-trivial modification of the diffeomorphism algebra.  

It was shown in \cite{Hohm:2014qga} how sphere compactifications of the original supergravities and their
non-compact cousins  can be realized in EFT through generalized Scherk-Schwarz 
compactifications, which are governed by E$_{d(d)}$ valued `twist' matrices. 
In terms of the duality covariant fields of EFT the reduction formulas  take the form 
of a simple Scherk-Schwarz ansatz (see (\ref{SSembedding}) below), 
proving the consistency of the corresponding Kaluza-Klein truncation. 
Although this settles the issue of consistency it 
may nevertheless be useful to have the explicit reduction formulas in terms of the 
\textit{conventional} supergravity fields. 
This requires the dictionary for identifying the original supergravity fields
in the EFT formulation.
In this paper we work out the explicit reduction formulas for 
the complete set of type IIB supergravity fields, using 
the general embedding of type IIB supergravity into the E$_{6(6)}$ EFT given in \cite{Baguet:2015xha}. 
In particular, this includes all components of the IIB self-dual four form.
Results for the scalar sector in the compact case have appeared in 
\cite{Khavaev:1998fb,Nastase:2000tu,Lee:2014mla,Ciceri:2014wya}.
The components of the twist matrix give rise to various conventional tensors, 
including for instance the Killing vectors in the case of $S^5$ but also various 
higher Killing-type tensors. We analyze the identities satisfied by these tensors 
by decomposing the Lie derivatives (\ref{LieIntro}), which can be thought of as 
giving generalized  Killing equations on the extended space. 
Various identities that appear miraculous from the point of view of standard geometry 
but are essential for consistency of the Kaluza-Klein ansatz are thereby 
explained in terms of the higher-dimensional E$_{6(6)}$ covariant geometry of EFT. 

This paper is not completely self-contained in that we assume some familiarity 
with the E$_{6(6)}$ EFT of \cite{Hohm:2013vpa}. Our recent review \cite{Baguet:2015xha}, which also gives the 
complete embedding of type IIB, can serve 
as a preparatory article. In particular, we use the same conventions. 
The rest of this paper is organized as follows.  
In sec.~2 we briefly review the generalized Scherk-Schwarz ansatz and the consistency conditions 
for the E$_{6(6)}$ EFT and give the twist matrices. 
The twist matrix gives rise to a set of generalized vectors of the extended space satisfying an 
algebra of generalized Lie derivatives (\ref{LieIntro}) akin to the algebra of Killing vector fields 
on a conventional manifold. In sec.~3 we analyze the various components of this equation 
and give the explicit solutions in terms of various Killing-type tensors. 
In sec.~4 we review the class of $D=5$ gauged supergravities that will be 
embedded into type IIB. Finally, in sec.~5 we work out the complete Kaluza-Klein ansatz 
by using the general embedding of type IIB established in \cite{Baguet:2015xha}. In particular, 
we show how to reconstruct the self-dual 4-form of type IIB from the EFT fields. 
Along the way, we show that the reduction ansatz reduces the ten-dimensional self-duality equations 
to the equations of motion of the $D=5$ theory. While this is guaranteed by the general argument,
its explicit realization requires an impressive interplay of Killing vector/tensor identities and the 
${\rm E}_{6(6)}/{\rm USp}(8)$ coset space structure of the five-dimensional scalar fields.
In sec.~6 we summarize the final results, the full set of reduction formulas,
and comment on the fermionic sector.
Some technically involved computations 
are relegated to an appendix.

\section{Generalized Scherk-Schwarz reduction}
We begin by giving the generalized Scherk-Schwarz ansatz in terms 
of the variables  of exceptional field theory.  This ansatz is governed 
by a group-valued twist matrix $U\in {\rm E}_{6(6)}$ and a 
scale factor $\rho$, both of which depend only on the internal coordinates $Y$. 
For the bosonic EFT fields, the general reduction ansatz reads~\cite{Hohm:2014qga}
\bea\label{SSembedding}
{\cal M}_{MN}(x,Y) &=& U_{M}{}^{\underline{K}}(Y)\,U_{N}{}^{\underline{L}}(Y)\,M_{\underline{K}\underline{L}}(x)\;, 
\nonumber\\
 g_{\mu\nu}(x,Y) &=& \rho^{-2}(Y)\,{\rm g}_{\mu\nu}(x)\;,\nonumber\\
  {\cal A}_{\mu}{}^{M}(x,Y) &=& \rho^{-1}(Y) A_{\mu}{}^{\underline{N}}(x)(U^{-1})_{\underline{N}}{}^{M}(Y) \;, 
  \nonumber\\
  {\cal B}_{\mu\nu\,M}(x,Y) &=& \,\rho^{-2}(Y) U_M{}^{\underline{N}}(Y)\,B_{\mu\nu\,\underline{N}}(x)
  \;.
 \eea
 Here, indices $M, N$ label the fundamental representation ${\bf 27}$ of ${\rm E}_{6(6)}$, and the four lines
 refer to the internal metric, external metric, vector fields and two-forms, respectively,
 see \cite{Hohm:2013vpa} for details.
In order for the ansatz (\ref{SSembedding}) to be consistent, $U$ and $\rho$ need to factor out 
homogeneously of all covariant expressions defining the action and equations 
of motion. This is the case provided the following two consistency equations (`twist equations') 
are satisfied:
 \bea\label{consistency}
   \partial_N(U^{-1})_{\underline{K}}{}^{N}-4\,(U^{-1})_{\underline{K}}{}^{N}\,\rho^{-1}\partial_N \rho 
  &=& 3\,\rho\,\vartheta_{\underline{K}}\;, 
\nonumber\\
  \big[(U^{-1})_{\underline{M}}{}^{K}(U^{-1})_{\underline{N}}{}^{L}\partial_{K}U_{L}{}^{\underline{P}}\big]_{\bf 351} 
  &=& \frac15\,\rho\,
  \Theta_{\underline{M}}{}^{{\boldsymbol{\alpha}}} t_{{\boldsymbol{\alpha}} \underline{N}}{}^{\underline{P}}\;.  
 \eea
Here the \textit{constant} tensors are  
$\vartheta_{\underline{K}}$, which defines the embedding tensor of `trombone' gaugings, 
and $\Theta_{\underline{M}}{}^{{\boldsymbol{\alpha}}}$, which defines the embedding tensor 
of conventional gaugings. 

For the subsequent analysis it is convenient to reformulate these
consistency conditions by rescaling the twist matrix by $\rho$, 
\bea
\widehat{U}^{-1} &\equiv& \rho^{-1}\,{U}^{-1}
\;.
\eea
This rescaling is such that $\widehat{U}^{-1}$ can be viewed as 
a generalized vector of the same density weight as the gauge parameters. 
Accordingly, one can define generalized Lie derivatives w.r.t.~this vector.  
The consistency conditions can then be brought into the compact form 
  \be\label{twistcompact}
  \mathbb{L}_{\,\widehat{U}^{-1}_{\underline{M}}}\,\widehat{U}^{-1}_{\underline{N}} \ \equiv  \ 
  -X_{\underline{M}\underline{N}}{}^{\underline{K}}\,\widehat{U}^{-1}_{\underline{K}}\;, 
 \ee
where $X_{\underline{M}\underline{N}}{}^{\underline{K}}$ are constants 
related to the $D=5$ embedding tensor by 
 \be\label{GaugeStructure}
  X_{\underline{M}\underline{N}}{}^{\underline{K}} \ = \ \left(\Theta_{\underline{M}}{}^{{\boldsymbol{\alpha}}}
  +\frac{9}{2}\,\vartheta_{\underline{L}}(t^{{\boldsymbol{\alpha}}})_{\underline{M}}{}^{\underline{L}}\right)
   (t_{{\boldsymbol{\alpha}}})_{\underline{N}}{}^{\underline{K}} - \delta_{\underline{N}}{}^{\underline{K}}\,\vartheta_{\underline{M}}\;. 
 \ee 
This implies in particular that the first equation in (\ref{consistency}) can be written as 
  \be\label{rhoCOND}
  \mathbb{L}_{\,\widehat{U}^{-1}_{\underline{M}}}\,\rho \ = \ -\vartheta_{\underline{M}}\,\rho\;. 
 \ee 
 
In \cite{Hohm:2014qga},  the consistency equations (\ref{consistency}) were solved for the 
sphere and hyperboloid compactifications, with gauge groups SO$(p,6-p)$ and CSO$(p,q,6-p-q)$,  
explicitly in terms of ${\rm SL}(6)$ group-valued twist matrices. 
Specifically, with the fundamental representation of ${\rm E}_{6(6)}$ decomposing as
\bea
\left\{Y^M\right\} &\longrightarrow& \{Y^{ab}, Y_{a\alpha}\}
\;,
\label{15120}
\eea  
into $(15,1)\oplus(6',2)$ under ${\rm SL}(6)\times {\rm SL}(2)$, we single out one of the 
fundamental ${\rm SL}(6)$ indices $a\rightarrow (0,i)$ to define the SL$(6)$ matrix $U_a{}^b$ as
\bea\label{twistfirst}
U_0{}^0 &\equiv&(1-v)^{-5/6}\left(1+u K(u,v)\right)\;,\nonumber\\
U_0{}^i &\equiv& -\eta_{ij}y^j \, (1-v)^{-1/3}\,K(u,v) \;,\nonumber\\
U_i{}^0 &\equiv& -\eta_{ij}y^j\,(1-v)^{-1/3}\, \;,\nonumber\\
U_i{}^j &\equiv& (1-v)^{1/6}\,\delta^{ij}
\;, 
\eea
with the combinations 
 \be\label{uvDEF} 
  u \ \equiv \  y^i\delta_{ij} y^j\;, \qquad v \ \equiv \  y^i\eta_{ij}y^j\;. 
 \ee 
Here $\eta_{ij}$ is the metric
\begin{align}
  \eta_{ij} &= 
  {\rm diag}\,(\,\underbrace{1, \dots,1,}_{p-1}\underbrace{-1,\dots,-1}_{6-p} 
  \,) 
\;, 
\label{etapqr}
\end{align}
and we define similarly the ${\rm SO}(p,6-p)$ invariant metric $\eta_{ab}$ with signature $(p,6-p)$.  
Note that in (\ref{uvDEF}) we use two different metrics, one Euclidean, the other pseudo-Euclidean. 
The function $K(u,v)$ is the solution of the differential equation 
\bea
2(1-v)\left(u \,\partial_v K+v\,\partial_u K \right)&=&
\left((7-2p)(1-v) -u\right) K -1
\;, 
\label{diffKsec2}
\eea
which can be solved analytically. For instance, for $p=6$, i.e., for gauge group SO(6) 
relevant for the S$^5$ compactification, the solution reads 
\bea
p=6 &:& \quad  K(u) \ = \  
\frac12\,u^{-3}\,\left(
u(u-3)+\sqrt{u(1-u)}\left(3\,{\rm arcsin}\sqrt{u}+c_0\right)
\right)
\;,
\eea
with constant $c_0$\,. We refer to \cite{Hohm:2014qga} for 
other explicit forms. 
The inverse twist matrix is given by
\bea
\label{Upq}
(U^{-1})_0{}^0&=&  (1-v)^{5/6}\;,
\nonumber\\
(U^{-1})_0{}^i &=& \eta_{ij} y^j\,(1-v)^{1/3}\,K(u,v)\;,
\nonumber\\
(U^{-1})_i{}^0 &=& \eta_{ij} y^j\,\,(1-v)^{1/3}\;,
\nonumber\\
(U^{-1})_i{}^j &=& (1-v)^{-1/6}\left( \delta^{ij}\,+  \eta_{ik} \eta_{jl} \, y^k  y^l\,K(u,v)\right)\;. 
\eea
Finally, the density factor $\rho$ is given by
\bea
\rho~=~
(1-v)^{1/6}
\;.
\label{rhopq}
\eea
 
Upon embedding the ${\rm SL}(6)$ twist matrix (\ref{twistfirst}) into ${\rm E}_{6(6)}$,
one may verify that it satisfies the consistency equations (\ref{consistency}) with an embedding
tensor that describes the gauge group ${\rm SO}(p,q)$, where the physical coordinates are embedded
into the EFT coordinates via (\ref{15120}) according to
\bea
y^i &=& Y^{[0i]} \;.
\eea

With the above form of the generalized Scherk-Schwarz ansatz and the explicit form 
of the twist matrix and the scale factor we have given the complete embedding of the 
corresponding sphere and hyperboloid compactifications into the E$_{6(6)}$ EFT. 
It is instructive, however, to clarify this embedding by analyzing it in terms of more conventional 
geometric objects. Therefore, in the next section we will analyze the consistency conditions  
(\ref{twistcompact}) under the appropriate decomposition (that embeds, for instance, 
the standard algebra of Killing vector fields on a sphere) and thereby reconstruct 
the above solution in a more conventional language. In particular, 
this will clarify the geometric significance of the function $K$, which is related 
to the four-form whose exterior derivative defines the volume form on the five-sphere.

\section{Untangling the twist equations}

\subsection{General analysis}
We now return to the `twist equations' (\ref{twistcompact}) and decompose them 
w.r.t.~the subgroup appropriate for the type IIB solution of the section constraint, i.e. 
\bea
{\rm E}_{6(6)} &\longrightarrow& {\rm GL}(5) \times {\rm SL}(2)
\;,
\nonumber\\
{\bf 27} &\longrightarrow& {\bf (5,1)} \oplus
{\bf (5',2)} \oplus {\bf (10,1)} \oplus{\bf (1,2)}
\;.
\label{GL5}
\eea
Accordingly, the fundamental index on the generalized vector $\widehat{U}^{-1}$ decomposes 
as 
\bea
(\widehat{U}^{-1})_{\underline{M}}{}^M &=&
\left\{\, {\cal K}_{\underline{M}}{}^m\,, \; {\cal R}_{\underline{M}\,m\alpha}\,,\;
{\cal Z}_{\underline{M}\, mnk}\,, \;{\cal S}_{\underline{M}\,n_1 \dots n_5 \alpha}
\,\right\}
\;,
\label{KRZS}
\eea
in terms of GL$(5)$ indices $m,n=1,\ldots,5$ and SL(2) indices $\alpha,\beta=1,2$. 
In order to give the decomposition of the twist equations (\ref{twistcompact}) in terms of these objects  
we use the definition (\ref{LieIntro}) of the generalized Lie derivative and the 
decomposition of the $d$-symbol (3.28) in \cite{Baguet:2015xha}. A straightforward computation, largely analogous 
to those in, e.g., sec.~3.3 of \cite{Baguet:2015xha}, then yields 
\bea\label{untangle1}
-X_{\underline{MN}}{}^{\underline{K}}\,
{\cal K}_{\underline{K}}{}^m &=& {\cal L}_{{\cal K}_{\underline{M}}} {\cal K}_{\underline{N}}{}^m\;,\\ \label{untangle2}
-X_{\underline{MN}}{}^{\underline{K}}\,
{\cal R}_{\underline{K}\,m\alpha} &=&
{\cal L}_{{\cal K}_{\underline{M}}} {\cal R}_{\underline{N}\,m\alpha}
-{\cal L}_{{\cal K}_{\underline{N}}} {\cal R}_{\underline{M}\,m\alpha}
+   \partial_m \left({\cal K}_{\underline{N}}{}^n {\cal R}_{\underline{M}\,n\alpha} \right)
\;, \\ \label{untangle3}
-X_{\underline{MN}}{}^{\underline{K}}\,
{\cal Z}_{\underline{K}\,kmn}  &=&
{\cal L}_{{\cal K}_{\underline{M}}} {\cal Z}_{\underline{N}\,kmn}
-{\cal L}_{{\cal K}_{\underline{N}}} {\cal Z}_{\underline{M}\,kmn}
+3\,  \partial_{[k}\left( {\cal K}_{\underline{N}}{}^l {\cal Z}_{\underline{M}\,mn]l}\right) 
\nonumber\\
&&{}
+3\sqrt{2} \, \varepsilon^{\alpha\beta} \, 
\partial_{[k} {\cal R}_{\underline{M}\,m|\alpha|} {\cal R}_{\underline{N}\,n]\beta}
\;,  \\
-X_{\underline{MN}}{}^{\underline{K}}\,
{\cal S}_{\underline{K}\,n_1 \dots n_5 \alpha} &=&
{\cal L}_{{\cal K}_{\underline{M}}} {\cal S}_{\underline{N}\,n_1 \dots n_5 \alpha}
\nonumber\\
&&{}
+20\sqrt{2}\left({\cal Z}_{\underline{N}\,[n_1n_2n_3} \partial_{n_4} {\cal R}_{\underline{M}\,n_5] \alpha}
- \partial_{[n_1}{\cal Z}_{\underline{M}\,n_2n_3n_4}  {\cal R}_{\underline{N}\,n_5] \alpha}
\right)
\;.
\label{untangle4}
\eea

We will now successively analyze these equations. 
We  split the index as $\underline{M}\rightarrow\{A,u\}$, 
where $A,B$ denote the `gauge group directions' and $u,v$ the remaining ones, 
and assume that
the only non-vanishing entries of $X_{\underline{MN}}{}^{\underline{K}}$ are
\bea
X_{AB}{}^C \ = \ - f_{AB}{}^C\;,\qquad
X_{Au}{}^v \ = \ (D_A)_u{}^v
\;,
\label{Au}
\eea
given in terms of structure constants and representation matrices of the underlying Lie algebra
of the gauge group, c.f.~\cite{deWit:2004nw}.
Let us emphasize that $X_{\underline{MN}}{}^{\underline{K}}$ is not assumed to be antisymmetric. 
In particular, for this ansatz we have, e.g., $X_{uA}{}^{v}=0$. 
Let us also stress that this ansatz  is not the most general, but it is sufficient for the
purposes in this paper. 

The first equation (\ref{untangle1}), specialized to external indices $(A,B)$, implies that the vector fields
${\cal K}_{A}$ satisfy the Lie bracket algebra 
\bea\label{KAlgebra}
\big[{\cal K}_{A},{\cal K}_{B}\big]^m \ \equiv \ {\cal L}_{{\cal K}_{A}} {\cal K}_{B}{}^m
&=& f_{AB}{}^C\,
{\cal K}_{C}{}^m\;. 
\eea
In view of standard Kaluza-Klein compactifications it is natural to interpret 
these vector fields as the Killing vectors of some internal geometry. 
We now \textit{define} a metric w.r.t.~which the ${\cal K}_{A}$ are indeed Killing vectors
by setting for the inverse metric 
\bea
\tilde{G}^{mn}  \ \equiv \ 
{\cal K}_A{}^m\, {\cal K}_B{}^n\,\eta^{AB} 
\;,
\label{Gtilde}
\eea
with  the Cartan-Killing metric $\eta_{AB}\equiv f_{AC}{}^D f_{BD}{}^C$.
The internal metric $\tilde{G}_{mn}$ exists provided the Cartan-Killing metric is invertible and 
that there are sufficiently many vector fields ${\cal K}_A{}^m$ to make $\tilde{G}^{mn}$
invertible. This assumption, which we will make throughout the following discussion, 
is satisfied in the examples below. Since by (\ref{KAlgebra}) the ${\cal K}_{A}$ 
transform under themselves according to the adjoint group action, under which the 
Cartan-Killing metric is invariant, it follows that the vectors are indeed Killing:
 \be\label{Killingeq}
  {\cal L}_{{\cal K}_{A}}\tilde{G}_{mn} \ \equiv \ \nabla_m{\cal K}_{An}+\nabla_n{\cal K}_{Am} \ = \ 0\;, 
 \ee
where here and in the following $\nabla_m$ denotes the covariant derivative 
w.r.t.~the metric (\ref{Gtilde}), which is used to raise and lower indices.  
The other non-trivial components of (\ref{untangle1}), with external indices $(A,u)$, $(u,A)$ 
and $(u,v)$, 
imply that the remaining vector fields ${\cal K}_{u}{}^m$ satisfy 
 \be
  {\cal L}_{{\cal K}_{A}}{\cal K}_{u}{}^m \ = \ -(D_{A})_{u}{}^{v}\,{\cal K}_{v}{}^m \ = \ 0\;, \qquad
  {\cal L}_{{\cal K}_{u}}{\cal K}_{v}{}^m \ \equiv \ \big[{\cal K}_{u}, {\cal K}_{v}\big]^m \ = \ 0\;. 
 \ee 
For non-vanishing ${\cal K}_{u}$ the first equation can only be satisfied if the representation 
encoded by the $(D_{A})_u{}^v$ includes the trivial (singlet) representation. 
In the following we will analyze the remaining equations under the assumption
that the representation does not contain a trivial part, which then requires   
\be\label{Kunull}
 {\cal K}_u{}^m \ = \ 0\;. 
\ee

We next consider the second equation (\ref{untangle2}), specialized to external indices $(A,u)$
and $(u,A)$ to obtain 
  \bea
{\cal L}_{{\cal K}_A} {\cal R}_{u\,m\alpha} &=&
-(D_A)_u{}^v\,{\cal R}_{v\,m\alpha}
~=~
\partial_m\left(
{\cal K}_{A}{}^n {\cal R}_{u\,n\alpha}
\right)
\;.
\label{calKR}
\eea
Writing out the Lie derivative on the left-hand side we obtain in particular 
\bea
{\cal K}_A{}^n \left(\partial_m {\cal R}_{u\,n\alpha} 
- \partial_n {\cal R}_{u\,m\alpha}\right) &=& 0
\;.
\eea
With the above assumption that the metric (\ref{Gtilde}) is invertible
it follows that the curl of ${\cal R}$ is zero. Hence we can write it in terms 
of a gradient, 
 \be\label{Rgradient}
 {\cal R}_{u\,m\alpha} \ \equiv \ \partial_m {\cal Y}_{u\,\alpha}\;. 
\ee 
As we still have to solve the first equation of (\ref{calKR}), we must demand that the function ${\cal Y}$ 
transforms under the Killing vectors in the representation $D_A$, 
\be\label{calKwY}
  {\cal L}_{{\cal K}_A}{\cal Y}_{u\,\alpha} \ = \ -(D_{A})_{u}{}^{v}\,{\cal Y}_{v\,\alpha}\;, 
 \ee 
for then (\ref{calKR}) follows with the covariant relation (\ref{Rgradient}). 
Finally, specializing (\ref{untangle2}) to external indices $(A,B)$, we obtain 
 \be
  f_{AB}{}^{C}{\cal R}_{C\,m\alpha} \ = \ {\cal L}_{{\cal K}_{A}}{\cal R}_{B\,m\alpha}
  -{\cal L}_{{\cal K}_{B}}{\cal R}_{A\,m\alpha}+\partial_m\big({\cal K}_{B}{}^n{\cal R}_{A\,n\alpha}\big)\;.
 \ee
This equation is solved by ${{\cal R}}_{A\,m\alpha}  =  0$, and the latter indeed holds for the  
${\rm SL}(6)$ valued twist matrix to be discussed below. 
In addition, we will find that for these twist matrices also the components ${\cal Z}_{u}$ and ${\cal S}_{A}$ 
are zero, and therefore in the following we analyze the equations for this special case, 
 \be\label{R=0}
   {{\cal R}}_{A\,m\alpha} \ = \ {\cal Z}_{u\,mnk} \ = \ {\cal S}_{A\,n_1\ldots n_5\,\alpha} \ = \ 0\;.  
 \ee  

Let us now turn to the third equation (\ref{untangle3}), which will constrain the  ${\cal Z}$ tensor. 
Specializing to external indices $(A,B)$, we obtain 
 \be
  f_{AB}{}^{C} \,{\cal Z}_{C\,kmn} \ = \ {\cal L}_{{\cal K}_{A}} {\cal Z}_{B\,kmn} 
  -{\cal L}_{{\cal K}_{B}} {\cal Z}_{A\,kmn} 
  +3\,\partial_{[k}\big({\cal K}_{B}{}^l {\cal Z}_{A\,mn]l}\big)\;,  
  \label{Zalgebra}
 \ee 
where we used (\ref{R=0}). Writing out the second Lie derivative 
on the right-hand side, this can be reorganized as 
\bea
{\cal L}_{{\cal K}_{A}} {\cal Z}_{B\,kmn} - 4\,{\cal K}_B{}^p\,\partial_{[p} {\cal Z}_{A\,kmn]}
&=& f_{AB}{}^{C}\,{\cal Z}_{C\,kmn}  
\;.
\label{equationZ}
\eea
In order to solve this equation we make the following ansatz 
\bea
 {\cal Z}_{A\,klm} &\equiv &
   -\frac14\,\sqrt{2}\,{\cal K}_{A\,klm} -2\,\sqrt{2} \,{\cal K}_{A}{}^p\, \tilde{C}{}_{pklm}
   \;, 
\label{defZ}
\eea
in terms of a four-form $\tilde{C}$,
where we chose the normalization for later convenience, 
and we defined the Killing tensor
\bea\label{KillingTensors}
{\cal K}_{A\,klm}&\equiv& \frac12\,\tilde{\omega}_{klmpq} \,{\cal K}_{A}{}^{pq}
\;,\qquad
{\cal K}_{A\,mn}~\equiv~ 2\,\nabla_{[m} {\cal K}_{A\,n]}
\;, 
\eea
with the volume form $\tilde{\omega}_{klmpq}\equiv |\tilde{G}|^{1/2}\, \varepsilon_{klmpq}$. 
We recall that all internal indices are raised and lowered with ${\tilde{G}}_{mn}$ defined in (\ref{Gtilde}).

It remains to determine $\tilde{C}{}_{pklm}$ from the above system of equations. 
In order to simplify the result of 
inserting (\ref{defZ}) into (\ref{equationZ}) we can use that the Killing tensor term 
transforms `covariantly' under the Lie derivative,
 \be
  {\cal L}_{{\cal K}_{A}}{\cal K}_{B\,mnk} \ = \ f_{AB}{}^{C}\,{\cal K}_{C\,mnk}\;, 
 \ee
which follows from the corresponding property (\ref{KAlgebra}) of the Killing vectors.   
For the second term on the left-hand side of (\ref{equationZ}), however, we have to compute, 
 \be\label{curlKcomp}
 \begin{split}
  {\cal K}_{B}{}^p\nabla_{[p}{\cal K}_{A\,kmn]} \ &= \ {\cal K}_{B}{}^p\nabla_{[p}\big(\,\tfrac{1}{2}\,\tilde{\omega}_{kmn]lq}
  \, {\cal K}_A{}^{lq} \,\big) \ = \ {\cal K}_{B}{}^p\, \tilde{\omega}_{lq[kmn}\nabla_{p]}\nabla^{[l}{\cal K}_{A}{}^{q]}\\
  \ &= \ -\tfrac{1}{2}\,{\cal K}_{B}{}^p\,\tilde{\omega}_{kmnpl}\,\nabla_{q}\nabla^{[l}{\cal K}_A{}^{q]}
  \ = \ \tfrac{1}{2}\,{\cal K}_{B}{}^p\,\tilde{\omega}_{kmnpl}\,\nabla_{q}\nabla^{q}{\cal K}_A{}^{l}\;. 
 \end{split}
 \ee 
Here we used the $D=5$ Schouten identity $\tilde{\omega}_{[lqkmn}\nabla_{p]}\equiv 0$
and that the Killing tensor written as 
${\cal K}_{A\,mn}=2\, \nabla_m{\cal K}_{A\,n}$ is automatically antisymmetric 
as a consequence of the Killing equations (\ref{Killingeq}).  
Using the latter fact again, the last expression simplifies as follows 
\bea\label{covCOMM}
\nabla_{q}
\nabla^q {\cal K}_{A}{}^l \ = \
-\nabla_{q}
\nabla^l {\cal K}_{A}{}^q
 \ = \ 
-\big[\nabla_{q},
\nabla^l \big]\, {\cal K}_{A}{}^q ~=~-\tilde{\cal R}^{lp}\,{\cal K}_{A\,p} 
\;.  
\eea
We will see momentarily that (\ref{equationZ}) can be solved analytically 
by the above ansatz (\ref{defZ}) if the metric $\tilde{G}$ is Einstein. 
We thus assume this to be the case, so that the Ricci tensor reads $\tilde{\cal R}_{mn}=\lambda\, \tilde{G}_{mn}$,
for some constant $\lambda$. Using this in (\ref{covCOMM}) 
and inserting back into (\ref{curlKcomp}) we obtain 
 \be\label{Result1}
   {\cal K}_{B}{}^p\nabla_{[p}{\cal K}_{A\,kmn]} \ = \ 
   \frac{\lambda}{2}\,\tilde{\omega}_{kmnpl}\,{\cal K}_{A}{}^p{\cal K}_{B}{}^l\;. 
 \ee
 Next, insertion of the second term in (\ref{defZ})  into (\ref{equationZ}) yields 
the contribution 
\be\label{Result2} 
{\cal L}_{{\cal K}_{A}} \big( {\cal K}_{B}{}^p\, \tilde{C}{}_{pkmn} \big) 
+ 4\,{\cal K}_B{}^p\,\partial_{[p} \big( {\cal K}_{A}{}^q\, \tilde{C}{}_{kmn]q} \big) \ = \
 f_{AB}{}^{C} {\cal K}_{C}{}^p\, \tilde{C}{}_{pkmn}
 +5\,{\cal K}_A{}^p {\cal K}_B{}^q\,\partial_{[p} \tilde{C}{}_{qkmn]}\;. 
\ee
Here we used (\ref{KAlgebra}) and combined the terms from ${\cal L}_{{\cal K}_A}\tilde{C}_{pkmn}$ 
with those from the second term on the left-hand side. 
Employing now (\ref{Result1}) and (\ref{Result2}) we find that insertion of (\ref{defZ}) into (\ref{equationZ}) 
yields 
 \be
   0 \ = \  {\cal K}_{A}{}^p {\cal
     K}_B{}^q\big(5\,\partial_{[p}\tilde{C}_{qkmn]}-\frac14\,\lambda\,\tilde{\omega}_{pqkmn}\big)\;.
\ee 
Thus, we have determined $\tilde{C}$, up to closed terms, to be  
 \be\label{dettildec}
  5\,\partial_{[p}\tilde{C}_{qkmn]} \ =  \ \frac14 \lambda\,\tilde{\omega}_{kmnpq}\;, 
 \ee 
which can be integrated to solve for $\tilde{C}_{klmn}$, 
since in five coordinates the 
integrability condition is trivially satisfied.
In total we have proved that the $(A,B)$ component of the third equation (\ref{untangle3})
of the system is solved by (\ref{defZ}).  
We also note that the remaining components of (\ref{untangle3}) are identically satisfied
under the assumption (\ref{R=0}). (For the $(u,v)$ component this requires 
using that the exterior derivative of ${\cal R}_{u\,m \alpha}$ vanishes by (\ref{Rgradient}).)  
For the subsequent analysis it will be important to determine how $\tilde{C}$ 
transforms under the Killing vectors. 
To this end we recall that in the 
definition (\ref{defZ}) $\tilde{C}$ is the only `non-covariant' contribution, which therefore accounts 
for the second term on the left-hand side of the defining equation 
(\ref{equationZ}). From this we read off 
 \be\label{calKc}
  {\cal L}_{{\cal K}_{A}}\tilde{C}_{mnkl} \ = \ -\,\sqrt{2}\,\partial_{[m}{\cal Z}_{A\,nkl]}\;. 
 \ee   

Finally, we turn to the last equation (\ref{untangle4}), 
which determines ${\cal S}_{u}$. Under the assumptions (\ref{Kunull}), (\ref{R=0}),  
the $(u,v)$ and $(u,A)$ components trivialize, while  
the $(A,u)$  component implies
 \be\label{SEQuation}
 \begin{split}
  {\cal L}_{{\cal K}_{A}}{\cal S}_{u\,n_1\ldots n_5\alpha} \ &= \
  -(D_A)_u{}^v {\cal S}_{v\,n_1\ldots n_5\alpha}
  +20\sqrt{2}\,\partial_{[n_1} {\cal Z}_{A\,n_2n_3n_4}\,{\cal R}_{|u| n_5]\alpha}\;. 
  \end{split}
 \ee 
We will now show that this equation is solved by 
 \be\label{Sansatz}
  {\cal S}_{u\,n_1\ldots n_5\alpha} \ = \ a\,\tilde{\omega}_{n_1\ldots n_5}\,{\cal Y}_{u\alpha}
  -20\,\tilde{C}_{[n_1\ldots n_4}\,\partial_{n_5]}{\cal Y}_{u\alpha}\;, 
 \ee
in terms of the volume form of $\tilde{G}_{mn}$, the function defined in (\ref{Rgradient})
and the four-form defined via (\ref{dettildec}). Here, $a$ is an arbitrary coefficient, while we 
set the second coefficient to the value 
that is implied by the following analysis. 
We first note that  ${\cal L}_{{\cal K}_A} \tilde{\omega}_{n_1\ldots n_5}  = 0$, 
which follows from the invariance under the Killing vectors of the metric $\tilde{G}$ defining 
$\tilde{\omega}$. 
Second, we recall (\ref{calKwY}), which states that the function  ${\cal Y}_{u}$
transforms `covariantly' under ${\cal L}_{{\cal K}_A}$ (i.e., w.r.t.~the representation matrices $D_A$). 
Thus, all terms in (\ref{Sansatz}) transform covariantly, except for 
the four-form $\tilde{C}$, whose `anomalous' transformation must therefore account for the 
second term in ${\cal L}_{{\cal K}_A}{\cal S}_u$ on the right-hand side of (\ref{SEQuation}). 
Using the anomalous transformations of  $\tilde{C}$ given in (\ref{calKc}), it then 
follows that (\ref{Sansatz}) solves (\ref{SEQuation})
for arbitrary coefficient $a$. 
This concludes our general discussion of the system of equations (\ref{untangle1})--(\ref{untangle4}).

\subsection{Explicit tensors}

We now return to the explicit twist matrices and read off the tensors whose general structure 
we discussed in the previous subsection. To this end we have to split the E$_{6(6)}$ indices further 
in order to make contact with the twist matrices given in (\ref{twistfirst}), (\ref{Upq}). 
As it turns out, for these twist matrices the split of indices $V_{\underline{M}} \ \equiv \ (V_{A},V_{u})$
discussed before (\ref{Au}), coincides with the split $27=15+12$ of (\ref{15120})
 \be
  V_{\underline{M}} \ \equiv \ (V_{A},V_{u}) \ \equiv \ (V_{[ab]}, V^{a\alpha})\;, \qquad
  a,b \ = \ 0,\ldots, 5\;, \quad \alpha, \beta \ = \ 1,2\;. 
  \label{1512}
 \ee 
In several explicit formulas we will have to split $[ab]$ further, 
 \be
  [ab] \ \equiv \ ([0i], [ij])\;, \qquad i,j=1\,\ldots, 5\;. 
 \ee  
Similarly, we perform the same index split for the fundamental index $M$ 
under E$_{6(6)}\rightarrow {\rm SL}(6)$ (and then further to ${\rm GL}(5) \times {\rm SL}(2)$
according to (\ref{GL5})), thus giving up in the following the distinction between bare and 
underlined indices. Let us note that we employ the convention 
 \be\label{convroot2}
  V^{0i} \ \equiv \ \tfrac{1}{\sqrt{2}}V^i\;, 
 \ee
in agreement with the summation conventions of ref.~\cite{Hohm:2013vpa}.  
In order to read off the various tensors from the twist matrices let us 
first canonically  embed the SL(6) matrix $U_{a}{}^{b}$ into E$_{6(6)}$. 
Under the above index split we have 
\bea
U_M{}^{\underline{N}} &=& 
 \left(
\begin{array}{cc}
U_{[ab]}{}^{[cd]}  & U_{[ab]}{}^{c\alpha}
\\[1ex]
U^{a\alpha,[cd]} & U^{a\alpha,}{}_{b\beta} 
\end{array}
\right)
\ = \ 
 \left(
\begin{array}{cc}
U_{[a}{}^{c}\,U_{b]}{}^{d}  & 0 
\\[1ex]
0 & \delta^\alpha{}_{\beta}\,(U^{-1})_{b}{}^{a}   
\end{array}
\right)
\;. 
\label{U6}
\eea

With this embedding, and recalling the convention (\ref{convroot2}), we can 
identify the Killing vector fields with components of 
the twist matrices as follows, 
 \be
  {\cal K}_{[ab]}{}^{m} \ \equiv \ \sqrt{2}\,(\widehat{U}^{-1})_{ab}{}^{m0}\;, 
 \ee 
which yields 
\bea
{\cal K}_{[0i]}{}^m(y) &=&- \frac12\,\sqrt{2}\,(1-v)^{1/2}\,\delta_i^m\;,\qquad
{\cal K}_{[ij]}{}^m(y) ~=~  \sqrt{2}\,\delta^m_{[i} \, \eta^{}_{j]k} y^k \;. 
\label{KV}
\eea
It is straightforward to verify that these vectors satisfy the Lie bracket 
algebra (\ref{KAlgebra}). 
Specifically, 
\bea
\big[\mathcal{K}_{ab},\mathcal{K}_{cd}\big]^{m} \ = \ -\sqrt{2} f_{ab\,,cd}{}^{ef}\mathcal{K}_{ef}{}^m\;,
\qquad
 f_{ab\,,cd}{}^{ef} \ \equiv \ 
 2\,\delta_{[a}{}^{[e}\eta_{b][c}\delta_{d]}{}^{f]}\;, 
 \label{fabc}
\eea
with the SO$(p,6-p)$ metric $\eta_{ab}$. The Killing tensors defined in (\ref{KillingTensors}) 
are then found to be 
 \be
 \begin{split}
  {\cal K}_{[0i]mnk} \ &= \ -\sqrt{2}\,\varepsilon_{mnkij}\,y^j\;, \\
  {\cal K}_{[ij]mnk} \ &= \ -\sqrt{2}(1-v)^{-\frac{1}{2}}\,\varepsilon_{mnkpq}
  \big(\,\delta_{i}{}^p\,\delta_{j}{}^q-2\,\delta_{[i}{}^p\eta_{j]l} \, y^q y^l\, \big)\;. 
 \end{split}
 \ee

We can now define the metric $\tilde{G}$ as in (\ref{Gtilde}) w.r.t.~which these 
vectors are Killing, using the Cartan-Killing form $\eta^{ab,cd}=\eta^{a[c}\eta^{d]b}$. 
This yields for the metric and its inverse 
\be
\begin{split}
\tilde{G}_{mn} \ &= \ \eta_{mn}+ (1-v)^{-1}\eta_{mp}\eta_{nq}y^py^q\;,\\
\tilde{G}^{mn} \ &= \ \eta^{mn}-y^my^n\;.
\end{split}
\ee
One may verify that this metric 
describes the homogeneous space ${\rm SO}(p,q)/{\rm SO}(p-1,q)$
with
\bea
\tilde{\cal R}_{mn} &=& 4 \,\tilde{G}_{mn}
\;, 
\eea
determining the constant above, $\lambda=4$. 
The associated volume form is given by  
 \be\label{volumeform}
  \tilde{\omega}_{mnklp} \ = \ (1-v)^{-\frac{1}{2}}\,\varepsilon_{mnklp} \;. 
 \ee

Next we give the function defining ${\cal R}$ in (\ref{Rgradient}) w.r.t.~the 
above index split, 
 \be\label{Rgradient2}
  {\cal R}_{u\,m\alpha} \ = \ {\cal R}^{a\beta}{}_{m\alpha} \ = \ \partial_m{\cal Y}^{a\beta}{}_{\alpha}\;, 
 \ee 
for which we read off from the twist matrix
\bea
 {\cal Y}^{a\beta}{}_{\alpha} \ = \ {\cal Y}^{a}\delta_{\alpha}^{\beta}\qquad\mbox{with}\qquad
 {\cal Y}^a(y) &\equiv& 
  \left\{
 \begin{array}{cc}
 (1-v)^{1/2} & a=0\\
 y^i & a=i
 \end{array}
 \right.
 \;.
 \label{defY}
 \eea
In agreement with (\ref{calKwY}) this transforms in the fundamental representation 
of the algebra of Killing vector fields (\ref{KV}). Specifically, 
\bea
{\cal L}_{{\cal K}_{[ab]}}{\cal Y}^c \ = \ 
{\cal K}_{[ab]}{}^m\,\partial_m {\cal Y}^c &=& \sqrt{2}\, \delta^c{}_{[a}\, {\cal Y}^{}_{b]}
\;, 
\label{idKY}
\eea
where ${\cal Y}_{a}$ is obtained from ${\cal Y}^a$ by means of $\eta_{ab}$. 
Let us also emphasize that the ${\cal Y}_{a}$ can be viewed as `fundamental harmonics', 
satisfying 
 \be
   \Box {\cal Y}^a \ = \ -5\, {\cal Y}^a \;, 
 \ee
in that all higher harmonics can then be constructed from them. For instance,  the Killing vectors themselves 
can be written as 
 \bea
  {\cal K}_{[ab]m} \ = \ \sqrt{2}\big(\partial_m{\cal Y}_{[a}\big)\, {\cal Y}_{b]}\;. 
 \label{idKYY}
 \eea

Next we compute the four-form $\tilde{C}_{mnkl}$ by integrating (\ref{dettildec}). 
An explicit solution can be written in terms of the function $K$ from (\ref{diffKsec2}) as 
\bea
\tilde{C}{}_{mnkl} &=&\frac{\lambda}{16}\,(1-v)^{-1/2} \,\varepsilon_{mnklq}\,
\big(K \delta^{qr}\eta_{rs}  + \delta_s^{q}\big)y^s
\;,
\label{Ctilde}
\eea
whose exterior derivative is indeed proportional to the volume form (\ref{volumeform}) for the metric $\tilde{G}_{mn}$. 
Together with 
the Killing vectors and tensors defined above, the ${\cal Z}$ tensor is now uniquely 
determined according to (\ref{defZ}). 
Moreover, it is related to the twist matrix according to 
 \be
  {\cal Z}_{[ab]mnk} \ = \ \frac{1}{2}\, \varepsilon_{mnkpq}\,\big(\widehat{U}^{-1}\big)_{[ab]}{}^{[pq]}
  \ = \ \frac{1}{2}\, \varepsilon_{mnkpq}\,\rho^{-1}\,(U^{-1})_{[a}{}^{p}\,(U^{-1})_{b]}{}^{q}\;, 
 \ee 
which agrees with (\ref{defZ}) for $\lambda=4$\,. 

Finally, let us turn to the tensor ${\cal S}_{u}$ whose general form is given in (\ref{Sansatz}).  
Under the above index split it is convenient  to write this tensor as 
 \be
   {\cal S}_{u\,n_1\ldots n_5\,\beta} \ \equiv \ {\cal S}^{a\alpha}{}_{n_1\ldots n_5\,\beta}
   \ \equiv \  {\cal S}^{a}\,\varepsilon_{n_1\ldots n_5}\,\delta^{\alpha}{}_{\beta}\;, 
 \ee  
which is read off from the twist matrix as 
 \be
   {\cal S}^{a\alpha}{}_{n_1\ldots n_5\,\beta} \ = \ \varepsilon_{n_1\ldots n_5}(\widehat{U}^{-1})^{a\alpha}{}_{0\beta}
   \ = \ \varepsilon_{n_1\ldots n_5}\,\rho^{-1}\,\delta^{\alpha}{}_{\beta}\, U_{0}{}^{a}\;, 
 \ee
leading with (\ref{twistfirst}) to    
 \bea
{\cal S}^a &=& 
 \left\{
 \begin{array}{cc}
 (1-v)^{-1}\,(1+uK) &\quad  a=0\\
\; -\eta_{ij}y^j\,(1-v)^{-1/2}\,K & \quad a=i
 \end{array}
 \right. 
 \;.
\eea 
One may verify that this agrees with (\ref{Sansatz}) for 
 \be
  a \ = \ 1\;, \qquad \lambda \ = \ 4\;. 
 \ee

\bigskip

\subsection{Useful identities}
\label{subsec:useful}

In this final paragraph we collect various identities satisfied by the above Killing-type 
tensors. These will be useful in the following sections when explicitly verifying the 
consistency of the Kaluza-Klein truncations. We find  
\bea
\mathcal{K}^{[ab]}{}_{mn}\mathcal{K}_{[cd]}{}^{n}
&=& 
-\sqrt{2}\, f_{cd,ef}{}^{ab}\mathcal{K}^{[ef]}{}_m
+2\, \partial_{m}\left( 
\delta_{[c}{}^{[a}\mathcal{Y}^{b]}\mathcal{Y}_{d]}
\right)\;,
\label{idKK2}
\\
{\cal K}^{[ab]}{}_n {\cal K}_{[cd]}{}^n &=&
2\,\delta_{[c}{}^{[a}\,\mathcal{Y}^{b]}\mathcal{Y}_{d]}
\;,
\label{idKK}
\\
{\cal K}_{[ab]}{}^k {\cal Z}_{[cd]\,kmn}
+{\cal K}_{[cd]}{}^k {\cal Z}_{[ab]\,kmn} &=&
-\frac18\,\varepsilon_{abcdef} \,{\cal K}^{[ef]}{}_{mn}
\label{idKZ1}
\;,
\\
{\cal K}^{[ab]}{}_{mn} {\cal K}_{[cd]}{}^m {\cal K}_{[ef]}{}^n &=&
4\,\sqrt{2}   \,\delta_{[c}{}^{[a} \, {\cal Y}_{d]}  {\cal Y}_{[e} \, \delta_{f]}{}^{b]}
\;,
\\
{\cal K}_{[cd]}{}^m {\cal K}_{[ab]}{}^n{\cal K}_{[ef]}{}^l\, \partial_l   {\cal K}^{[ab]}{}_{mn}  
&=&
-8\,\eta_{e[c}\,{\cal Y}_{d]} {\cal Y}_{f}  
+8\,\eta_{f[c}\,{\cal Y}_{d]} {\cal Y}_{e}  
\;, 
\eea
which can be verified using the explicit tensors determined above.

\newpage

\section{The $D=5$ supergravity}

The $D=5$ gauged theory with gauge group ${\rm SO}(p,q)$ was originally constructed in~\cite{Gunaydin:1984qu,Pernici:1985ju,Gunaydin:1985cu}.
For our purpose, the most convenient description is its covariant form found in the context of general 
gaugings~\cite{deWit:2004nw} to which we refer for details.\footnote{
To be precise, and to facilitate the embedding of this theory into EFT, 
we choose the normalization of \cite{Hohm:2013vpa} for vector and tensor fields which differs from \cite{deWit:2004nw}
as
\bea
{\cal A}_\m{}^M{}_{[1312.0614]}= \frac1{\sqrt{2}}{A}_\m{}^M{}_{[{\rm hep-th}\slash0412173]}\;,\quad
{\cal B}_{\m\n\, M}{}_{[1312.0614]}=-\frac14{B}_{\m\n\, M}{}_{[{\rm hep-th}\slash0412173]}\;,
\label{AABB}
\eea
 together with a rescaling of the associated symmetry parameters. Moreover, we have set the coupling constant 
 of \cite{deWit:2004nw} to $g=1$\,.
}
In the covariant formulation, the $D=5$ gauged theory features 27 propagating vector fields $A_\mu{}^M$ and up to 
$27$ topological tensor fields $B_{\mu\nu\,M}$. The choice of gauge group and the precise number of tensor fields
involved is specified by the choice of an embedding tensor $Z^{MN}=Z^{[MN]}$ in the 
${\bf 351}$ representation of ${\rm E}_{6(6)}$. E.g.\ the full non-abelian vector field strengths are given by
\bea
 F_{\mu\nu}{}^M &=& 2 \, \partial_{[\mu} A_{\nu]}{}^M 
+ \sqrt{2}\,X_{KL}{}^M A_{[\mu}{}^K A_{\nu]}{}^L -2\,\sqrt{2}\,Z^{MN} B_{\mu\nu\,N}
\;,
\label{FAB}
\eea
with the tensor $X_{KL}{}^M$ carrying the gauge group structure constants and defined 
in terms of the embedding tensor $Z^{MN}$ as
\bea
X_{MN}{}^P &=& d_{MNQ} Z^{PQ} + 10\,d_{MQS}d_{NRT}d^{PQR}Z^{ST}
\;.
\label{XMNP}
\eea

The ${\rm SO}(p,q)$ gaugings preserve the global ${\rm SL}(2)$ subgroup of the 
symmetry group ${\rm E}_{6(6)}$ of the ungauged theory, more specifically the centralizer of
its subgroup ${\rm SL}(6)$\,. Accordingly, the vector fields in the ${\bf 27}$ of ${\rm E}_{6(6)}$
can be split as
\bea
A_\mu{}^M &\longrightarrow&
\left\{ A_\mu{}^{ab}, A_{\mu\,a\alpha} \right\}
\;,\qquad
a, b  = 0, \dots, 5\;,\quad \alpha=1,2
\;,
\label{splitA}
\eea
into 15 ${\rm SL}(2)$ singlets and 6 ${\rm SL}(2)$ doublets, c.f.~(\ref{1512}).
The 27 two-forms $B_{\mu\nu\,M}$ split accordingly, with only the 6 ${\rm SL}(2)$ doublets $B_{\mu\nu}{}^{a\alpha}$
entering the supergravity Lagrangian.
In the basis (\ref{splitA}), the only non-vanishing components of
the embedding tensor $Z^{MN}$ are
\bea
Z_{a\alpha,b\beta} &\equiv& -\frac12\,\sqrt{5}\,\varepsilon_{\alpha\beta}\eta_{ab}
\;,
\eea
where the normalization has been chosen such as to match the later expressions.
With (\ref{XMNP}), we thus obtain\footnote{
The totally symmetric cubic $d$-symbol of ${\rm E}_{6(6)}$ in 
the ${\rm SL}(6)\times {\rm SL}(2)$ basis (\ref{splitA}) is given by
 \bea
d^{MNK} &:&
d^{ab}{}_{c\alpha,d\beta} = \frac1{\sqrt{5}}\, \delta^{ab}_{cd}\,\varepsilon_{\alpha\beta}\;,\quad
d^{ab,cd,ef} = \frac1{\sqrt{80}}\,\varepsilon^{abcdef}\;.
\eea
}
\bea
X_{MN}{}^K &:& 
\left\{
\begin{array}{l}
X_{ab,cd}{}^{ef} =
f_{ab,cd}{}^{ef}\\[1ex]
 X_{ab}{}^{c\alpha}{}_{d\beta} =
-\delta_{[a}{}^{c}  \eta_{b]d}\,\delta^\alpha_\beta 
 \end{array}
 \right.
 \;,
 \label{XXX}
\eea
with the ${\rm SO}(p,6-p)$ structure constants
$f_{ab,cd}{}^{ef}$ from (\ref{fabc}).

The form of the field strength (\ref{FAB}) is the generic structure of a covariant field strength in
gauged supergravity, with non-abelian Yang-Mills part and a St\"uckelberg type coupling to the two-forms.
In the present case, we can make use of the tensor gauge symmetry which acts by shift
$\delta A_{\mu\,a\alpha}=\Xi_{\mu\,a\alpha}$ on the vector fields, to eliminate all components $A_{\mu\,a\alpha}$
from the Lagrangian and field equations. This is the gauge we are going to impose in the following,
which brings the theory in the form of~\cite{Gunaydin:1985cu}.\footnote{
To be precise: this holds with a rescaling of $p$-forms according to
\bea
{\cal A}_\m{}^{ab}{}_{[1312.0614]}=-\sqrt{2}\,A_\mu{}^{ab}{}_{\rm GRW}\;,\qquad
\sqrt{5} \,{\cal B}_{\mu\nu}{}^{a\alpha}{}_{[1312.0614]}=B_{\mu\nu}{}^{a\alpha}{}_{\rm GRW}\;, 
\label{scaleGRW}
\eea
and with their coupling constant set to $g_{\rm GRW}=2$\,.}
As a result, the covariant object (\ref{FAB}) splits into components carrying the ${\rm SO}(p,q)$ Yang-Mills field strength,
and the two-forms $B_{\mu\nu}{}^{a\alpha}$, respectively, 
\bea
 F_{\mu\nu}{}^M &=&
 \left\{
 \begin{array}{rcl}
 F_{\mu\nu}{}^{ab} &\equiv&  2 \, \partial_{[\mu} A_{\nu]}{}^{ab} 
+ \sqrt{2}\,f_{cd,ef}{}^{ab}\,A_\mu{}^{cd} A_\nu{}^{ef}  
\\[1ex]
F_{\mu\nu\,a\alpha} &\equiv&  
\sqrt{10}\,\varepsilon_{\alpha\beta}\eta_{ab}\,B_{\mu\nu}{}^{b\beta}
\end{array}
\right.
\;.
\label{F12}
\eea
In particular, fixing of the tensor gauge symmetry
implies that the two-forms $B_{\mu\nu}{}^{a\alpha}$ turn into topologically massive fields, preserving the correct
counting of degrees of freedom, \cite{Townsend:1983xs}. The Lagrangian and field equations  
are still conveniently expressed in terms of the combined object $F_{\mu\nu}{}^M$.
E.g.\ the first order duality equation between vector and tensor fields is given by
\bea
3\, D_{[\mu} B_{\nu\rho]}{}^{a\alpha}  &=&\frac{1}{2\sqrt{10}}  \,\sqrt{|{\rm g}|}\, \varepsilon_{\mu\nu\rho\sigma\tau} 
{M}^{a\alpha}{}_{N}\,{ F}^{\sigma\tau\,N}
\;,
\label{duality5}
\eea
which upon expanding around the scalar origin and with (\ref{F12}) yields the first order topologically massive field 
equation for the two-form tensors.
The full bosonic Lagrangian reads
\bea
{\cal L}&=&  \sqrt{|{\rm g}|}\,R 
   - \frac1{4}\,\sqrt{|{\rm g}|}\,M_{MN}\,F_{\mu\nu}{}^M F^{\mu\nu\,N}
  +\frac1{24} \,\sqrt{|{\rm g}|}\,D_\mu M_{MN} D^\mu M^{MN}   
    \nn\\[1ex] 
&&{}
+  \varepsilon^{\m\n\rho\sigma\tau} \left(
\frac54\, \varepsilon_{\alpha\beta}\,\eta_{ab}\,
 B_{\m\n}{}^{a\alpha}  D_\rho B_{\sigma\tau}{}^{b\beta} + \frac1{24}\, \sqrt{2}\,   \varepsilon_{abcdef} \, A_\mu{}^{ab}\,\partial_\nu 
   A_\rho{}^{cd} \,\partial_\sigma A_\tau{}^{ef} \right)
   \nn\\
&&{}
+ \frac1{16}\,\varepsilon^{\m\n\rho\sigma\tau}  \varepsilon_{abcdef} \,  f_{gh,ij}{}^{ab} \, A_\m{}^{cd} A_\n{}^{gh} A_{\rho}{}^{ij}
 \Big(\partial_\sigma  A_{\tau}{}^{ef}+\ft15 \sqrt{2}\, 
 f_{kl,mn}{}^{ef} A_\sigma{}^{kl}A_\tau{}^{mn} \Big) \nonumber \\[1ex]
&&{} 
  - \sqrt{|{\rm g}|}\,V(M_{MN})
 \,.
 \label{LD5}
\eea
Here, the 42 scalar fields parameterize the coset space ${\rm E}_{6(6)}/{\rm USp}(8)$
via the symmetric ${\rm E}_{6(6)}$ matrix $M_{MN}$ which can be decomposed in the basis (\ref{splitA})
as
\bea
M_{MN} &=& \left(
\begin{array}{cc}
{M}_{ab,cd}&{M}_{ab}{}^{c\gamma}\\
{\cal M}^{a\alpha}{}_{bc} & M^{a\alpha,c\gamma}
\end{array}
\right)
\;,
\label{MD5}
\eea
with the ${\rm SO}(p,6-p)$ covariant derivatives defined according to
\bea
D_\mu X^a &\equiv& \partial_\mu X^a + \sqrt{2}\,A_\mu{}^{ab} \, \eta_{bd}\, X^d
\;,
\label{covSO}
\eea
and similarly on the different blocks of (\ref{MD5}). 
The scalar potential $V$ in (\ref{LD5}) is given by the following
contraction of the generalized structure constants (\ref{XXX}) with the scalar matrix (\ref{MD5})
\bea
V(M_{MN}) &=&
\frac{1}{30}\,M^{MN}X_{MP}{}^Q
\left(5\, X_{NQ}{}^P + X_{NR}{}^S \,M^{PR}M_{QS}
\right)\;.
\label{potential}
\eea

For later use, let us explicitly state the vector field equations obtained from (\ref{LD5})
which take the form
\bea
0 &=&
 \sqrt{|{\rm g}|}\,\varepsilon_{\mu\nu\rho\sigma\tau}
\left( \eta_{c[a} \,D^{\tau} M_{b]d,N} M^{N,cd}
+\sqrt{2}\,
D_\lambda\left( {F}^{\tau\lambda\,N} {M}_{N,ab}\right) \right)
\nonumber\\
&&{}
+
\frac{3}{2}\,\,
\varepsilon_{abcdef} \,F_{[\mu\nu}{}^{cd} F_{\rho\sigma]}{}^{ef}
 +60 \,
 \varepsilon_{\alpha\beta} \,\eta_{ac}\eta_{bd}\,
 B_{[\mu\nu}{}^{c\alpha}  B_{\rho\sigma]}{}^{d\beta} 
 \;.
 \label{eomV}
\eea
We will also need part of the scalar field equations 
that are obtained by varying in (\ref{LD5}) the scalar matrix (\ref{MD5}) with 
an ${\rm SL}(6)$ generator $X_a{}^b$
\bea
0 &=&  \frac1{4} \,D^\mu (M^{ad\,K}    D_\mu M_{K\,bd} )  
   - \frac1{2}\, M_{bc\,N} \,F_{\mu\nu}{}^{ac} F^{\mu\nu\,N}
 + \frac1{4}\,\sqrt{10}\,\eta_{bc}\,\varepsilon_{\alpha\beta}\,
 M^{a\alpha}{}_{N} \,B_{\mu\nu}{}^{c\beta} F^{\mu\nu\,N}
\nonumber\\
&&{}
+\left(2\,M^{ae,fc}
+ \frac{4}{15}\, {M}^{de,h(a}  {M}^{c)j,fg} {M}_{dg,hj}    
+ \frac1{15}\,{M}^{de,h(a}  {M}^{c)\beta,f\alpha} {M}_{d\alpha,h\beta}   \right) \eta_{bc}\eta_{ef} 
\nonumber\\
&&{}
- \frac{2}{15}\left(
{M}^{de,k(a}  M^{c)\alpha}{}_{dg} {M}_{k\alpha}{}^{fg} 
+ {M}^{de,h(a}     {M}^{c)g}{}_{d\alpha} {M}^{f\alpha}{}_{hg}  
\right)
\eta_{ef} \eta_{bc}  
~-~[{\rm trace}]_b{}^a
\;.\qquad
\label{D5scalarEOM}
\eea

\section{The IIB reduction ansatz}

In terms of the ${\rm E}_{6(6)}$ EFT fields, the reduction ansatz is given by the 
simple factorization (\ref{SSembedding}) with the twist matrix $U$ given by (\ref{Upq}). 
In order to translate this into the original IIB theory,
we may first decompose the EFT fields under (\ref{GL5}),
according to the IIB solution of the section constraint, and collect the expressions
for the various components. We do this separately for EFT vectors, two-forms,
metric, and scalars, 
and subsequently derive the expressions for three- and four-forms from the IIB self-duality equations,
as outlined in the general case in \cite{Baguet:2015xha}.
In a second step, we can then recombine the various EFT components into the original IIB fields,
upon applying the explicit dictionary~\cite{Hohm:2013vpa,Baguet:2015xha} from IIB into EFT.

In particular, the explicit expression for the full IIB metric allows one to determine the
background metric, i.e.\ the IIB metric at the point where all $D=5$ scalar fields are set to zero.
This metric may or may not extend to a solution of the IIB field equations, depending on whether 
the scalar potential of the $D=5$ theory has a stationary point at its origin.
It is known~\cite{Gunaydin:1985cu} that this is the case for the $D=5$
theories with gauge group ${\rm SO}(6)$ and ${\rm SO}(3,3)$, with AdS and dS vacuum, respectively.
Accordingly, the internal manifolds $S^5$ and $H^{3,3}$ extend to
solutions of the full IIB field equations, with the external geometry given by
${\rm AdS}_5$ or ${\rm dS}_5$, respectively.

\subsection{IIB supergravity}

Let us briefly review our conventions for the $D=10$ IIB supergravity~\cite{Schwarz:1983wa,Schwarz:1983qr,Howe:1983sra}. 
The IIB field equations can be most compactly obtained from the pseudo-action
\bea
S &=& \int d^{10}\hat{x} \, \sqrt{|G|}\;\Big(\hat{R}+\frac{1}{4} \partial_{\hat{\mu}}m_{\alpha \beta}\partial^{\hat{\mu}}m^{\alpha \beta}-\frac{1}{12}\hat{F}_{\hat{\mu}_1\hat{\mu}_2\hat{\mu}_3}{}^{\alpha}\hat{F}^{\hat{\mu}_1\hat{\mu}_2\hat{\mu}_3}{}^{\beta}m_{\alpha \beta}\nonumber\\
&&{}
\qquad\qquad\qquad\qquad
-\frac{1}{30}\hat{F}_{\hat{\mu}_1\hat{\mu}_2\hat{\mu}_3\hat{\mu}_4\hat{\mu}_5}\hat{F}^{\hat{\mu}_1\hat{\mu}_2\hat{\mu}_3\hat{\mu}_4\hat{\mu}_5}\Big)\nonumber\\
&&{}-\frac{1}{864}\,
\int d^{10}\hat{x}\,\varepsilon_{\alpha\beta}\,\varepsilon^{\hat{\mu}_1\ldots \hat{\mu}_{10}}C_{\hat{\mu}_1\hat{\mu}_2\hat{\mu}_3\hat{\mu}_4}\hat{F}_{\hat{\mu}_6\hat{\mu}_7\hat{\mu}_8}{}^{\alpha}\hat{F}_{\hat{\mu}_8\hat{\mu}_9\hat{\mu}_{10}}{}^{\beta}
\label{actionIIB}
\;.
\eea
Here, $D=10$ coordinates are denoted by $x^{\hat\mu}$, and the action carries the field strengths
\bea
  \hat{F}_{\hat{\mu}\hat{\nu}\hat{\rho}}{}^{\alpha} &\equiv&
  3\,\partial_{[\hat{\mu}}\hat{C}_{\hat{\nu}\hat{\rho}]}{}^{\alpha}\;,
  \nonumber\\
   \hat{F}_{\hat{\mu}_1\ldots\hat{\mu}_5} &\equiv& 5\,\partial_{[\hat{\mu}_1}\hat{C}_{\hat{\mu}_2\ldots \hat{\mu}_5]}
  -\frac54 \, 
  \varepsilon_{\alpha\beta}\,\hat{C}_{[\hat{\mu}_1\hat{\mu}_2}{}^{\alpha}\hat{F}_{\hat{\mu}_3\hat{\mu}_4\hat{\mu}_5]}{}^{\beta}\;,
\label{FFIIB}
\eea
of two- and four-form gauge potential.
After variation, the field equations derived from (\ref{actionIIB}) have to be supplemented with the standard 
self-duality equations for the 5-form field strength 
\bea
\hat{F}_{\hat{\mu}\hat{\nu}\hat{\rho}\hat\sigma\hat\tau}&=&
\frac1{5!}\,\sqrt{|G|}\,\varepsilon_{\hat{\mu}\hat{\nu}\hat{\rho}\hat\sigma\hat\tau
\hat\mu_1\hat\mu_2\hat\mu_3\hat\mu_4\hat\mu_5}\,
\hat{F}^{\hat\mu_1\hat\mu_2\hat\mu_3\hat\mu_4\hat\mu_5}
\;.
\label{FF55}
\eea
Finally, the symmetric ${\rm SL}(2)$ matrix $m_{\alpha\beta}$ parametrizes the coset space 
${\rm SL}(2)/{\rm SO}(2)$ and carries dilaton and axion. In the notation of~\cite{Schwarz:1983qr}
it is parametrized by a complex scalar $B$ as
\bea
m_{\alpha\beta} &\equiv& (1-BB^*)^{-1} \left(
\begin{array}{cc}
(1-B)(1-B^*) & i (B-B^*) \\
 i (B-B^*)  & (1+B)(1+B^*)
 \end{array}
\right)
\;.
\eea
As a first step for the reduction ansatz, we perform the $5+5$ Kaluza-Klein
decomposition of coordinates $\{x^{\hat\mu}\}=\{x^\mu, y^m\}$ and fields, 
starting from the ten-dimensional vielbein
\be
E_{\hat{\mu}}{}^{\hat{a}}=\left(\begin{array}{cc}
({\rm det}\,\phi)^{-1/3}\;e_{\mu}{}^{\underline{a}} & 
A_{\mu}{}^{m}\phi_{m}{}^{\underline{\alpha}} \\
 0  & \phi_{m}{}^{\underline{\alpha}}
 \end{array}\right)\;, 
 \label{E10}
 \ee
but keeping the dependence on all 10 coordinates. Decomposition of the $p$-forms
in standard Kaluza-Klein manner then involves the projector 
$P_{\mu}{}^{\hat{\nu}}=E_{\mu}{}^{\underline{a}} E_{\underline{a}}{}^{\hat{\nu}}$ 
together with a further redefinition of fields due to the Chern-Simons contribution in (\ref{FFIIB}), 
see~\cite{Baguet:2015xha} for details. This leads to the components 
\bea
{C}_{mn}{}^{\alpha} &\!\equiv\!& \hat{C}_{mn}{}^{\alpha}\;,\nonumber\\
{C}_{\mu\, m}{}^{\alpha} &\!\equiv\!& \hat{C}_{\mu m}{}^{\alpha}-A_{\mu}{}^p\hat{C}_{pm}{}^{\alpha}\;,\nonumber\\
{C}_{\mu\nu}{}^{\alpha} &\!\equiv\!& \hat{C}_{\mu \nu}{}^{\alpha}-2A_{[\mu}{}^{p}\hat{C}_{|p|\nu]}{}^{\alpha}+A_{\mu}{}^{p}A_{\nu}{}^{q}\hat{C}_{pq}{}^{\alpha}\;,\nonumber\\[1ex]
{C}_{mnkl} &\!\equiv\!& \hat{C}_{mnkl}\;,\nonumber\\
{C}_{\mu \, nkl} &\!\equiv\!& \hat{C}_{\mu nkl}-A_{\mu}{}^{p}\hat{C}_{pnkl}
-\ft38\varepsilon_{\alpha \beta}\,{C}_{\mu \,[n}{}^{\alpha}{C}_{kl]}{}^{\beta}
\;,\nonumber\\
{C}_{\mu \nu\, kl}&\!\equiv\!& \hat{C}_{\mu \nu kl}-2A_{[\mu}{}^{p}\hat{C}_{|p|\nu]kl}+A_{\mu}{}^{p}A_{\nu}{}^{q}\hat{C}_{pqkl}
-\ft18\,\varepsilon_{\alpha \beta}\,{C}_{\mu\nu}{}^{\alpha}{C}_{kl}{}^{\beta}\;,\nonumber\\
{C}_{\mu \nu \rho \,m}&\!\equiv\!& \hat{C}_{\mu \nu \rho\, m}-3A_{[\mu}{}^{p}\hat{C}_{|p|\nu \rho]\,m}+3A_{[\mu}{}^{p}A_{\nu}{}^{q}\hat{C}_{|pq|\rho ]\,m}-A_{\mu}{}^{p}A_{\nu}{}^{q}A_{\rho}{}^{r}\hat{C}_{pqrm}
\nonumber\\
&&{}
-\ft38\varepsilon_{\alpha \beta}
\,{C}_{[\mu \nu}{}^{\alpha}{C}_{\rho]\,m}{}^{\beta}
\;,\nonumber\\
{C}_{\mu \nu \rho \sigma}&\!\equiv\!& \hat{C}_{\mu \nu \rho \sigma}-4A_{[\mu}{}^{p}\hat{C}_{|p|\nu \rho \sigma]}+6A_{[\mu}{}^{p}A_{\nu}{}^{q}\hat{C}_{|pq|\rho \sigma]}
-4A_{[\mu}{}^{p}A_{\nu}{}^{q}A_{\rho}{}^{r}\hat{C}_{|pqr| \sigma]}
\nonumber\\
&&{}
+A_{\mu}{}^{p}A_{\nu}{}^{q}A_{\rho}{}^{r}A_{\sigma}{}^{s}\hat{C}_{pqrs}
\;.
\label{C2C4}
\eea
in terms of which the
reduction ansatz is most naturally given in the following.

\subsection{Vector and two-form fields}

Breaking the 27 EFT vector fields according to (\ref{GL5}) into
\bea
\left\{ {\cal A}_\mu{}^m, {\cal A}_{\mu\,m\alpha},{\cal A}_{\mu\,kmn}, {\cal A}_{\mu\,\alpha}\right\}
\;,
\label{A27}
\eea
we read off the reduction ansatz from (\ref{SSembedding}), (\ref{KRZS}), 
which in particular gives rise to
\bea
{\cal A}_\mu{}^{m}(x,y) &=& {\cal K}_{[ab]}{}^m(y)\,A_\mu^{ab}(x)
\;,
\nonumber\\
{\cal A}_{\mu\,kmn}(x,y) &=& {\cal Z}_{[ab]\,kmn}(y)\,A_\mu^{ab}(x)
\;. 
\label{AA1}
\eea
The Kaluza-Klein vector field ${\cal A}_\mu{}^{m}={A}_\mu{}^{m}$
thus reduces in the standard way with the 15 Killing vectors ${\cal K}_{[ab]}{}^m(y)$
whose algebra defines the gauge group of the $D=5$ theory.
Note, however, that these extend to Killing vectors of the internal space-time metric 
only in case of the compact gauge group ${\rm SO}(6)$.
In the general case, as discussed above, the ${\cal K}_{[ab]}{}^m(y)$ are the Killing vector fields of 
an auxiliary homogeneous Lorentzian metric (\ref{Gtilde}), compare 
also~\cite{Hull:1988jw,Cvetic:2004km,Baron:2014bya}.
The vector field components ${\cal A}_{\mu\,kmn}$ are expressed in terms of the same 15 
$D=5$ vector fields. Their internal coordinate dependence is not exclusively carried by Killing vectors and tensors,
but exhibits via the tensor ${\cal Z}_{[ab]\,kmn}(y)$ an inhomogeneous term carrying the 
four-form $\tilde{C}{}_{mnkl}$ according to (\ref{defZ}).\footnote{
This seems to differ from the ansatz derived in \cite{Ciceri:2014wya}. The precise comparison should
take into account that the $A_\mu$, $B_{\mu\nu}$ are non-gauge-invariant vector potentials. In the present discussion, 
the inhomogeneous term in ${\cal Z}_{[ab]\,kmn}(y)$
played a crucial role in the verification of the proper algebraic relations.}
This is similar to reduction formulas for the dual vector fields in the $S^7$ reduction 
of $D=11$ supergravity~\cite{Godazgar:2013pfa}, which, however, in the present case 
already show up among the fundamental vectors.

For the remaining vector field components, the ansatz (\ref{SSembedding}), (\ref{KRZS}),
at first yields the reduction formulas
\bea
{\cal A}_{\mu\,m\alpha}(x,y) &=& 
{\cal R}^{a\beta}{}_{m\alpha}(y) \, A_{\mu a\beta}(x) \ = \ 
\partial_m {\cal Y}^a(y)\,A_{\mu\,a\alpha}(x)\;,
\nonumber\\
{\cal A}_{\mu\,\alpha}(x,y) &=& {\cal S}^{a}(y) \, A_{\mu\,a\alpha}(x) \\ \nonumber 
& = & |\tilde{G}|^{1/2}  \left( \, {\cal Y}^{a}(y) 
  -\frac16\,\tilde\omega^{klmnp} \tilde{C}_{klmn}\,\partial_{p}{\cal Y}^a(y)\right)A_{\mu\,a\alpha}(x)\;,
\label{AA2}
\eea
in terms of the 12 vector fields $A_{\mu\,a\alpha}$ in $D=5$ and the tensors defined in (\ref{Sansatz}) 
and (\ref{Rgradient2}).
However, as discussed in the previous section, for the ${\rm SO}(p,q)$ gauged theories,
a natural gauge fixing of the two-form tensor gauge transformations allows to eliminate 
these vector fields in exchange for giving topological mass to the two-forms.
As a result, the final reduction ansatz reduces to
\bea
{\cal A}_{\mu\,m\alpha}  \ = \ 0 \ = \  
{\cal A}_{\mu\,\alpha} \;.
\label{AA3}
\eea

\addtocounter{footnote}{-1}

For the two-forms, upon breaking them into ${\rm GL}(5)$ components
\bea
\left\{ {\cal B}_{\mu\nu}{}^\alpha, {\cal B}_{\mu\nu\,mn}, {\cal B}_{\mu\nu}{}^{m\alpha}, {\cal B}_{\mu\nu\,m} \right\}
\;,
\eea
similar reasoning via (\ref{SSembedding}) and evaluation of the twist matrix $\rho^{-2}\,U_M{}^{\underline{N}}$
gives the following ansatz for the ${\rm SL}(2)$ doublets
\bea
{\cal B}_{\mu\nu}{}^\alpha(x,y) &=& {\cal Y}_a(y)\,B_{\mu\nu}{}^{a\alpha}(x)
\;,\nonumber\\
{\cal B}_{\mu\nu}{}^{m\alpha}(x,y) &=&
{\cal Z}_a{}^m(y)\,B_{\mu\nu}{}^{a\alpha}(x)
\;,
\label{BB1}
\eea
in terms of the 12 topologically massive two-form fields of the $D=5$ theory.
Here, ${\cal Z}_a{}^m(y)$ is the vector density, given by\footnotemark
\bea
{\cal Z}_a{}^m &=& 
|\tilde{G}|^{1/2}\left(
\tilde{G}^{mn} \partial_n {\cal Y}_a
+\frac1{6}\,
\tilde\omega^{mklpq} \tilde{C}_{klpq} {\cal Y}_a
\right)
\;,
\label{ZA}
\eea
in terms of the Lorentzian metric $\tilde{G}_{mn}$, vector field ${\cal Y}_a$, and four-form $\tilde{C}_{klmn}$.
As is obvious from their index structure, the fields ${\cal B}_{\mu\nu}{}^{m\alpha}$
contribute to the dual six-form doublet of the IIB theory, but not to the original IIB fields.
Accordingly, for matching the EFT Lagrangian to the IIB dynamics, these fields are
integrated out from the theory~\cite{Hohm:2013vpa,Baguet:2015xha}.
For the IIB embedding of $D=5$ supergravity, we will thus only need the 
first line of (\ref{BB1}).

For the remaining two-form fields, the reduction ansatz (\ref{SSembedding})
yields the explicit expressions
\bea
{\cal B}_{\mu\nu\,m}(x,y) &=&
{\cal Z}^{[ab]}{}_m(y) \,B_{\mu\nu\,ab}(x)
\;, \nonumber\\
{\cal B}_{\mu\nu\,mn}(x,y) &=&-\frac14\,\sqrt{2}\,{\cal K}^{[ab]}{}_{mn}(y) \,B_{\mu\nu\,ab}(x)
\;,
\label{BB2}
\eea
with the Killing tensor ${\cal K}^{[ab]}{}_{mn}=2\,\partial_{[m} {\cal K}^{[ab]}{}_{n]}$, and the tensor 
density ${\cal Z}^{[ab]}{}_m$
given by
\bea
{\cal Z}^{[ab]}{}_m &=&  |\tilde{G}|^{1/2}\left(
 {\cal K}^{[ab]}{}_m
+\frac1{12}\,\tilde\omega^{klnpq}
{\cal K}^{[ab]}{}_{mk}\,\tilde{C}_{lnpq}\right)
\;.
\eea
Here, the 15 $D=5$ two-forms $B_{\mu\nu\,ab}$ are in fact absent in the ${\rm SO}(p,q)$ supergravities,
described in the previous section.
In principle, they may be introduced on-shell, employing 
the formulation of these theories given in~\cite{deWit:2004nw,deWit:2008ta},  
however, subject to an additional (three-form tensor) gauge freedom, 
which subsequently allows one to set them to zero. Hence, in the following we adopt $B_{\mu\nu\,ab}(x)=0$, such that
(\ref{BB2}) reduces to
\bea
{\cal B}_{\mu\nu\,m}  \ = \ 
0 \ = \  {\cal B}_{\mu\nu\,mn} 
\;.
\label{BB3}
\eea
Within EFT, consistency of this choice with the reduction ansatz (\ref{BB2}) can be understood by the fact that the fields
${\cal B}_{\mu\nu\,m}$ (related to the IIB dual graviton) 
do not even enter the EFT Lagrangian, while the fields ${\cal B}_{\mu\nu\,mn}$
enter subject to gauge freedom
\bea
\delta {\cal B}_{\mu\nu\,mn} &=& 2\,\partial_{[m} \Lambda_{n]\,\mu\nu}
\;,
\eea
(descending from tensor gauge transformations of the IIB four-form potential), 
which allows us to explicitly gauge the reduction ansatz (\ref{BB2}) to zero.

Combining the reduction formulas for the EFT fields
with the explicit dictionary given in sec.~5.2 of~\cite{Baguet:2015xha}, we can use the results of this section 
to give the explicit expressions for the different components (\ref{C2C4}) of the type IIB form fields.
This gives the following reduction formulae
\bea
{C}_{\mu \nu}{}^{\alpha}(x,y) &=&  \sqrt{10}\, {\cal Y}_a(y)\,B_{\mu\nu}{}^{a\alpha}(x)
\;,\\ \nonumber 
{C}_{\mu m}{}^{\alpha}(x,y) &=&  0\;,
\nonumber\\
C_{\mu \nu \,mn}(x,y) &=&  \frac{\sqrt{2}}{4}\,
{\cal K}_{[ab]}{}^k(y) {\cal Z}_{[cd]\,kmn}(y)\,
{A}_{[\mu}{}^{ab}(x) {A}_{\nu]}{}^{cd}(x)\;,
\quad
\nonumber\\
C_{\mu kmn}(x,y) &=&  \frac{\sqrt{2}}{4}\,\mathcal{Z}_{[ab] \,kmn}(y)\,A_\mu{}^{ab}(x)
\;,\qquad
\label{formsIIB}
\eea
for two- and four form gauge potential in the basis after standard Kaluza-Klein
decomposition.
In the next subsection, we collect the expressions for the 
scalar components $C_{mn}{}^\alpha$ and $C_{klmn}$, and in 
subsection~\ref{subsec:34form} we derive the reduction formulas
for the last missing components $C_{\mu\nu\rho\,m}$, and $C_{\mu\nu\rho\sigma}$
of the four-form.

Let us finally note that with the reduction formulas given in this section,
also the non-abelian EFT field strengths of the vector fields factorize
canonically, as can be explicitly verified
with the identities given in (\ref{KAlgebra}), (\ref{Zalgebra}).
Explicitly, we find
 \bea
 \mathcal{F}_{\mu \nu}{}^m &\equiv& 
 2\,\partial_{[\mu} {\cal A}_{\nu]}{}^m 
- {\cal A}_{\mu}{}^n\partial_n {\cal A}_{\nu}{}^m
+{\cal A}_{\nu}{}^n\partial_n {\cal A}_{\mu}{}^m
\nonumber\\
&=&
 \mathcal{K}_{[ab]}{}^m(y)\left(
 2\, \partial_{[\mu}A_{\nu]}{}^{ab}(x)+\sqrt{2}\, f_{cd\,,ef}{}^{ab}A_{\mu}{}^{cd}A_{\nu}{}^{ef}(x) \right) 
 \nonumber\\
&=&
  \mathcal{K}_{[ab]}{}^m(y)\,F_{\mu\nu}{}^{ab}(x)
 \;,
 \nonumber\\[1ex]
  {\cal F}_{\mu\nu\,kmn} &\equiv&
2\,\partial_{[\mu} {\cal A}_{\nu]\,kmn}
-2\,{\cal A}_{[\mu}{}^l \partial_l {\cal A}_{\nu]\,kmn}
-3\,\partial_{[k} {\cal A}_{[\mu}{}^{l}  {\cal A}_{\nu]\,mn]l} 
+3\, {\cal A}_{[\mu}{}^{l}  \partial_{[k}{\cal A}_{\nu]\,mn]l}
\nonumber\\
&=& {\cal Z}_{[ab]\,kmn}(y)\,F_{\mu\nu}{}^{ab}(x)
\;,
\label{FKK}
\eea
in terms of  the non-abelian ${\rm SO}(p,q)$ field strength  $F_{\mu\nu}{}^{ab}(x)$
from (\ref{F12}).

\subsection{EFT scalar fields and metric}

Similar to the discussion of the form fields, the reduction of the EFT scalars
can be read off from (\ref{SSembedding}) upon proper parametrization of the matrix ${\cal M}_{MN}$.
We recall from \cite{Hohm:2013vpa,Baguet:2015xha} that ${\cal M}_{MN}$ is a real symmetric ${\rm E}_{6(6)}$
matrix parametrized by the 42 scalar fields
\bea
\left\{ G_{mn}, C_{mn}{}^\alpha, C_{klmn}, m_{\alpha\beta} \right\}
\;,  
\label{scalars}
\eea
where $C_{mn}{}^\alpha=C_{[mn]}{}^\alpha$, and $C_{klmn}=C_{[klmn]}$ are fully antisymmetric
in their internal indices, $G_{mn}=G_{(mn)}$ is the symmetric $5\times5$ matrix, representing
the internal part of the IIB metric, and  $m_{\alpha\beta}=m_{(\alpha\beta)}$ is the unimodular 
symmetric $2\times2$ matrix parametrizing the coset space ${\rm SL}(2)/{\rm SO}(2)$ 
carrying the IIB dilaton and axion.
Decomposing the matrix ${\cal M}_{MN}$ into blocks according to the basis (\ref{A27})
\bea
{\cal M}_{KM} &=& \left(
\begin{array}{cccc}
{\cal M}_{k,m}&{\cal M}_k{}^{m\beta}&{\cal M}_{k,mn}&{\cal M}_{k}{}^{\beta}\\
{\cal M}^{k\alpha}{}_{m}&{\cal M}^{k\alpha,}{}^{m\beta}&{\cal M}^{k\alpha}{}_{mn}&{\cal M}^{k\alpha,}{}^{\beta}\\
{\cal M}_{kl,m}&{\cal M}_{kl}{}^{m\beta}&{\cal M}_{kl,mn}&{\cal M}_{kl}{}^{\beta}\\
{\cal M}^\alpha{}_{m}&{\cal M}^{\alpha,m\beta}&{\cal M}^\alpha{}_{mn}&{\cal M}^{\alpha,\beta}
\end{array}
\right)
\;,
\label{M2B}
\eea
the scalar fields (\ref{scalars}) can be read off from the various components of
${\cal M}_{MN}$ and its inverse ${\cal M}^{MN}$\,.
We refer to \cite{Baguet:2015xha} for the explicit formulas and collect the final result
\bea
G^{mn}&=&
({\rm det} \,G)^{1/3}\, {\cal M}^{m,n}
  \;,
 \label{scalarsinM} \\
 m^{\alpha\beta} &=&  ({\rm det} \,G)^{2/3} {\cal M}^{\alpha,\beta}
  \;,\nonumber\\
C_{mn}{}^\alpha
 &=&
 \sqrt{2}\,  \varepsilon^{\alpha\beta}  ({\rm det} \,G)^{2/3}\,m_{\beta\gamma} \,{\cal M}^{\gamma}{}_{mn} 
 ~=~
-\varepsilon^{\alpha\beta} ({\rm det} \,G)^{1/3} \,G_{nk} \, {\cal M}_{m\beta}{}^k
  \;,\nonumber\\
C_{klmn} &=& \frac18\, ({\rm det} \,G)^{2/3}\,\varepsilon_{klmnp}\,m_{\alpha\beta} \,{\cal M}^{\alpha,p\beta}
~=~
-\frac{\sqrt{2}}{16}\,({\rm det} \,G)^{1/3} \varepsilon_{klmnp} G_{qr}\,{\cal M}^{pq,r}
 \;,
 \nonumber
 \eea
where $G^{mn}$ and $m^{\alpha\beta}$ denote the inverse matrices of $G_{mn}$ and $m_{\alpha\beta}$
from (\ref{scalars}). The last two lines represent examples how the $C_{mn}{}^\alpha$ and $C_{klmn}$ can be obtained
in different but equivalent ways either from components of ${\cal M}_{MN}$ or ${\cal M}^{MN}$.
This of course does not come as a surprise but is a simple 
consequence of the fact that the $27\times27$ matrix ${\cal M}_{MN}$ 
representing the 42-dimensional coset space ${\rm E}_{6(6)}/{\rm USp}(8)$ is subject to a large number
of non-linear identities.

With (\ref{scalarsinM}), the reduction formulas for the EFT scalars are immediately derived from (\ref{SSembedding}).
For the IIB metric and dilaton/axion, this gives rise to the expressions
\bea
 G^{mn}(x,y) &=& \,\Delta^{2/3}(x,y)\, {\cal K}_{[ab]}{}^m(y) {\cal K}_{[cd]}{}^n(y)\,M^{ab,cd}(x)
\;,\nonumber\\
m^{\alpha\beta}(x,y) &=&  
 \Delta^{4/3}(x,y)\, {\cal Y}_a(y) {\cal Y}_b(y) \, M^{a\alpha,b\beta}(x)
\;,
\label{reductionGm}
\eea
with the function $\Delta(x,y)$ defined by
\bea
\Delta(x,y) ~\equiv~  \rho^3(y)\,({\rm det}\, G)^{1/2}~=~(1-v)^{1/2}\,({\rm det}\, G)^{1/2}\;,
\eea
and the 42 five-dimensional scalar fields parameterizing the symmetric ${\rm E}_{6(6)}$ matrix $M_{MN}$
decomposed into an ${\rm SL}(6)\times {\rm SL}(2)$ basis as (\ref{MD5}).

Similarly, the reduction formula for the internal components of the two-form $C_{mn}{}^\alpha$ 
is read off as
\bea
C_{mn}{}^\alpha(x,y)
 &=&
-\varepsilon^{\alpha\beta}   \Delta^{2/3}(x,y) \, G_{nk}(x,y)\,
\partial_m {\cal Y}^c(y)\, {\cal K}_{[ab]}{}^{k}(y)\,M^{ab}{}_{c\beta}(x)  
\nonumber\\
&=&
-\frac12\,  
\varepsilon^{\alpha\beta} \Delta^{4/3}(x,y)\, m_{\beta\gamma}(x,y)\, {\cal Y}_c(y)\,{\cal K}^{[ab]}{}_{mn}(y)
\,{M}_{ab}{}^{c\gamma}(x)
\;,
\label{reductionb}
\eea
featuring the inverse matrices of (\ref{reductionGm}),
with the two alternative expressions corresponding to using the different equivalent
expressions in (\ref{scalarsinM}).
To explicitly show the second equality in (\ref{reductionb}) requires rather non-trivial
quadratic identities among the components (\ref{MD5}) of an ${\rm E}_{6(6)}$ matrix, together with
non-trivial identities among the Killing vectors and tensors.
In contrast, this identity simply follows on general grounds from the equivalence of the
two expressions in (\ref{scalarsinM}), i.e., it follows from the group property of ${\cal M}_{MN}$ and the
twist matrix $U_M{}^{\underline{N}}$.
Let us also stress, that throughout all indices on the Killing vectors ${\cal K}_{[ab]}{}^m$ and tensors are raised and
lowered with the Lorentzian $x$-independent metric $\tilde{G}_{mn}(y)$ from (\ref{Gtilde}), not with the
space-time metric $G_{mn}(x,y)$.

Eventually, the same reasoning gives the reduction formula for $C_{mnkl}$
\bea
C_{klmn}(x,y) &=& \frac18\,\varepsilon_{klmnp}\,   \Delta^{4/3}(x,y) \,m_{\alpha\beta}(x,y) \,
{\cal Y}_a(y)\,  {\cal Z}_b{}^p(y) \, M^{a\alpha,b\beta}(x)
\;,
\eea
with ${\cal Z}_b{}^p(y)$ from (\ref{ZA}). Explicitly, this takes the form
\bea
C_{klmn}(x,y) &=& 
\frac1{16} \,\tilde\omega_{klmnp}\,   \Delta^{4/3}(x,y) \,m_{\alpha\beta}(x,y) \,
 \tilde{G}^{pq}(y) \, \partial_q \left(\Delta^{-4/3}(x,y) \,m^{\alpha\beta}(x,y)\right)
 \nonumber\\
&&{}
+ \tilde{C}_{klmn}(y) 
 \;.
 \label{reductionC1}
\eea
On the other hand, using the last identity in (\ref{scalarsinM}) to express $C_{klmn}$,
the reduction formula is read off as 
\bea
C_{klmn}(x,y) &=& 
\frac{\sqrt{2}}{4}\,
\Delta^{2/3}(x,y)
\,{\cal Z}_{[ab]\,[klm}(y)  G_{n]r}(x,y)\, {\cal K}_{[cd]}{}^r(y) \, M^{ab,cd}(x)
\nonumber\\
&=&
\tilde{C}_{mnkl}(y) \, 
-\frac18\,
\Delta^{2/3}(x,y)
\,{\cal K}_{[ab]}{}^p(y) \,{\cal K}_{[cd]\,[klm}(y)  \, G_{n]p}(x,y)\, M^{ab,cd}(x)
\;,
\label{reductionC2}
\eea
where we have used the explicit expression (\ref{defZ}) for ${\cal Z}_{[ab]\,klm}$.
Again, the equivalence between (\ref{reductionC1}) and (\ref{reductionC2}) is
far from obvious, but a consequence of the group property of ${\cal M}_{MN}$ and the
twist matrix $U_M{}^{\underline{N}}$.
For the case of the sphere $S^5$, several of these reduction formulas have appeared
in the literature~\cite{Khavaev:1998fb,Nastase:2000tu,Pilch:2000ue,Lee:2014mla,Ciceri:2014wya}. Here we find that
they naturally generalize to the case of hyperboloids, inducing the $D=5$
non-compact ${\rm SO}(p,q)$ gaugings.

Let us finally spell out the reduction ansatz for the five-dimensional metric 
which follows directly from (\ref{SSembedding}) as
\bea
g_{\mu\nu}(x,Y) &=& \rho^{-2}(y)\,{\rm g}_{\mu\nu}(x)
\;.
\eea
Putting this together with the parametrization of the IIB metric in terms of the EFT fields,
and the reduction (\ref{AA1}) of the Kaluza-Klein vector field,
we arrive at the full expression for the IIB metric
\bea
ds^2 &=&
 \Delta^{-2/3}(x,y)\,{\rm g}_{\mu\nu}(x) \, dx^\mu dx^\nu  
\nonumber\\
 &&{}
 + 
G_{mn}(x,y)
\left( dy^m + {\cal K}_{[ab]}{}^m(y) A_\mu^{ab}(x)  dx^\mu\right)\left( dy^n +{\cal K}_{[cd]}{}^n(y) A_\nu^{cd}(x)   dx^\nu\right)
 \;,
 \label{metricIIB}
\eea
in standard Kaluza-Klein form \cite{Duff:1986hr}, with $G_{mn}$ given by the inverse of (\ref{reductionGm}).

\subsection{Background geometry}

It is instructive to evaluate the above formulas at the particular point where all $D=5$ fields vanish,
i.e.\ in particular the scalar matrix $M_{MN}$ reduces to the identity matrix
\bea
M_{MN}(x) &=& \delta_{MN}
\;.
\label{M=I}
\eea
This determines the background geometry around which the generalized Scherk-Schwarz reduction ansatz
captures the fluctuations.
Depending on whether or not the scalar potential of $D=5$ gauged supergravity has a stationary point at the origin ---
which is the case for the ${\rm SO}(6)$ and ${\rm SO}(3,3)$ gaugings~\cite{Gunaydin:1985cu} ---
this background geometry will correspond to a solution of the IIB field equations.

With (\ref{M=I}) and the vanishing of the Kaluza-Klein vector fields, 
the IIB metric (\ref{metricIIB}) reduces to
\bea
ds^2 &=& \overset{\circ}{G}{}_{\hat\mu\hat\nu}\,dX^{\hat\mu} dX^{\hat\nu} 
 \label{IIBbackground}
\\
&\equiv&
 (1+u-v)^{1/2}\,\overset{\circ}{{g}}_{\mu\nu}(x) \, dx^\mu dx^\nu  
 +  (1+u-v)^{-1/2}\left( \delta_{mn} + \frac{\eta_{mi}\eta_{nj} y^i y^j}{1-v} \right)
dy^m dy^n \;,
\nonumber
\eea
where we have used the relations
\bea
\delta^{ac}\delta^{bd}\,{\cal K}_{[ab]}{}^m(y) {\cal K}_{[cd]}{}^n(y) &=& (1+u-v)\,\delta^{mn} - \eta_{mi}\eta_{nj} y^i y^j
\;,\nonumber\\
\overset{\circ}{\Delta} &=& (1+u-v)^{-3/4}
\;.
\eea
The internal metric of (\ref{IIBbackground}) is conformally equivalent to the hyperboloid
$H^{p,6-p}$ defined by the embedding of the surface 
\bea
z_1^2 + \dots + z_p^2 - z_{p+1}^2 - \dots - z_6^2 &\equiv& 1
\;,
\eea
in $\mathbb{R}^6$\,. This is a Euclidean five-dimensional space with isometry group
${\rm SO}(p)\times {\rm SO}(6-p)$, inhomogeneous for $p=2,3,4$.
Except for $p=6$, this metric differs from the homogeneous Lorentzian metric defined in (\ref{Gtilde})
with respect to which the Killing vectors and tensors parametrizing the reduction ansatz are defined.

Using that ${\cal Y}_a {\cal Y}_b \, \delta^{ab} = 1+u-v$,
it follows from (\ref{reductionGm}) that the IIB dilaton and axion are constant
 \bea
 \overset{\circ}{m}{}^{\alpha\beta} &=& \delta^{\alpha\beta}
 \;,
 \eea
while the internal two-form (\ref{reductionb}) vanishes due to the fact that (\ref{M=I}) does not break the ${\rm SL}(2)$\,.
Eventually, the four-form $C_{klmn}$ is most conveniently evaluated from (\ref{reductionC1}) as
\bea
\overset{\circ}{C}{}_{klmn} &=& 
\tilde{C}_{klmn}-
\frac16\, \tilde{\omega}_{klmnp}\,  \tilde{G}^{pq}\,  
\overset{\circ}{\Delta}{}^{-1}  \partial_q \overset{\circ}{\Delta} 
 \nonumber\\
 &=&
 \frac14\, {\varepsilon}_{klmnp}\,\eta^{pq} y^q\,
  (1-v)^{-1/2} \, \left(K(u,v)+(1+u-v)^{-1}\right)  
 \;,
 \label{c_klmn}
\eea
which can also be confirmed from (\ref{reductionC2}).
In particular, its field strength is given by
\bea
5\,\partial_{[k}\overset{\circ}{C}{}_{lmnp]} &=&
\frac12\,\varepsilon_{klmnp}\,\frac{p-4+ (p-3)(u-v)}{(1-v)^{1/2}\,(1+u-v)^2}\;,
\label{dc0}
\eea
where we have used the differential equation (\ref{diffKsec2}) for the function $K(u,v)$.
Again, it is only for $p=6$, that the background four-form potential $\overset{\circ}{C}{}_{klmn}$ coincides with
the four-form $\tilde{C}_{klmn}$ that parametrizes the twist matrix $U_M{}^{\underline{N}}$\,.

With this ansatz, the type IIB field equations reduce to the Einstein equations, which in this normalization take the form
\bea
\overset{\circ}{R}_{mn} &=& \overset{\circ}{T}_{mn} ~\, \equiv\, ~
\frac{25}{6}\,\partial_{[m}\overset{\circ}{C}{}_{klpq]}\partial_{[n}\overset{\circ}{C}{}_{rstu]}\,
\overset{\circ}{G}{}^{kr}\overset{\circ}{G}{}^{ls}\overset{\circ}{G}{}^{pt}\overset{\circ}{G}{}^{qu}
\;,
\eea
and similar for $\overset{\circ}{R}_{\mu\nu}$\,. With (\ref{IIBbackground}) and (\ref{dc0}), the energy-momentum tensor
takes a particularly simple form for $p=6$ and $p=3$:
\bea
\overset{\circ}{T}_{mn} &=&
\left\{
\begin{array}{ll}
4\,
\overset{\circ}{G}{}_{mn} &  p=6\\
(1+u-v)^{-5/2}\,\overset{\circ}{G}{}_{mn} & p=3
\end{array}
\right.
\;.
\eea
For the $x$-dependent background metric $\overset{\circ}{{g}}_{\mu\nu}(x)$
the most symmetric ansatz assumes an Einstein space (dS, AdS, or Minkowski)
\bea
R[\overset{\circ}{g}]_{\mu\nu} &=& k\,\overset{\circ}{g}_{\mu\nu}
\;,
\label{EinK}
\eea
upon which the IIB Ricci tensor associated with (\ref{IIBbackground}) 
turns out to be blockwise proportional to the IIB metric for the same two cases $p=6$ and $p=3$
\bea
\overset{\circ}{R}_{mn} &=&
\left\{
\begin{array}{ll}
4\,
\overset{\circ}{G}{}_{mn} &  p=6\\
(1+u-v)^{-5/2}\,\overset{\circ}{G}{}_{mn} & p=3
\end{array}
\right.
\;,
\nonumber\\
\overset{\circ}{R}_{\mu\nu} &=&
\left\{
\begin{array}{ll}
k\,
\overset{\circ}{G}{}_{\mu\nu} &  p=6\\
-(1+u-v)^{-5/2}\left(
1+(2-k)\,(1+u-v)^2
\right)
\overset{\circ}{G}{}_{\mu\nu} & p=3
\end{array}
\right.
\;.
\eea
Together it follows that (\ref{IIBbackground}), (\ref{c_klmn}), (\ref{EinK}) 
solve the IIB field equations for $p=3, k=2$ and $p=6, k=-4$, c.f.~\cite{Hull:1988jw}.
The resulting backgrounds are AdS$_5\times S^5$ and dS$_5\times H^{3,3}$ and the
induced $D=5$ theories correspond to the ${\rm SO}(6)$ and the ${\rm SO}(3,3)$ gaugings of \cite{Gunaydin:1985cu},
respectively.
For $3\not=p\not=6$, the background geometry is not a solution to the IIB field equations. Let us stress, however, that
also in these cases the reduction ansatz presented in the previous sections describes a consistent truncation of the
IIB theory to an effectively $D=5$ supergravity theory, but this theory does not have a simple ground state with all fields vanishing.

\subsection{Reconstructing 3-form and 4-form}
\label{subsec:34form}

We have in the previous sections derived the reduction formulas for all EFT scalars, vectors, and two-forms.
Upon using the explicit dictionary into the IIB fields~\cite{Hohm:2013vpa,Baguet:2015xha}, this allows to reconstruct 
the major part of the original IIB fields. More precisely, among the components of the fundamental IIB fields only 
$\hat{C}_{\mu\nu\rho\,m}$ and $\hat{C}_{\mu\nu\rho\sigma}$ with three and four external legs of the IIB four-form 
potential remain undetermined from the previous analysis. These in turn can be reconstructed from 
the IIB self-duality equations, which are induced by the EFT dynamics. We refer to~\cite{Baguet:2015xha} for 
the details of the general procedure, which we work out in the following with the generalized Scherk-Schwarz
reduction ansatz.

The starting point is the duality equation between EFT vectors and two-forms that follows from the Lagrangian 
\bea
\partial_{[k}\left(
\tilde{\cal H}_{|\mu\nu\rho|\,mn]}-\frac12\,e\,
{\cal M}_{mn],N}\, {\cal F}^{\sigma\tau\,N} \,\varepsilon_{\mu\nu\rho\sigma\tau} \right) &=& 0
\;, 
\label{dualHF}
\eea
where ${\cal F}_{\mu\nu}{}^{N}$ is the non-abelian field strength associated with the vector fields ${\cal A}_\mu{}^N$,
and $\tilde{\cal H}_{|\mu\nu\rho|\,mn}$ carries the field strength of the two-forms ${\cal B}_{\mu\nu\,mn}$.
Taking into account the reduction ansatz (\ref{AA3}), (\ref{BB3}), it takes the explicit form
\bea
\tilde{\cal H}_{\mu\nu\rho\,mn} &=&
-\partial_{[\mu} {\cal A}_{\nu}{}^{k}  {\cal A}_{\rho]\,kmn}
- {\cal A}_{[\mu}{}^{k} \partial_\nu  {\cal A}_{\rho]\,kmn}
- {\cal F}_{[\mu\nu}{}^{k}  {\cal A}_{\rho]\,kmn}
-  {\cal A}_{[\mu}{}^{k} {\cal F}_{\nu\rho]\,kmn}
\nonumber\\
&&{}
+2\, \partial_{[m}\left( {\cal A}_{[\mu}{}^{k}  {\cal A}_{\vphantom{[}\nu}{}^l {\cal A}_{\rho]\,n]kl} \right)
\;,
\label{Htilde}
\eea
in terms of the remaining vector fields and field strengths from (\ref{FKK}).
Since (\ref{dualHF}) is of the form of a vanishing curl, the equation can be integrated in the
internal coordinates up to a curl $\partial_{[m} C_{n]\,\mu\nu\rho}$ related to the 
corresponding component of the IIB four-form, explicitly
\bea
\partial_{[m} C_{n]\,\mu\nu\rho} &=& 
\frac1{16}\sqrt{2}\,e\,\varepsilon_{\mu\nu\rho\sigma\tau}\,
{\cal M}_{mn,N}\, {\cal F}^{\sigma\tau\,N} 
- \frac18\sqrt{2}\,\tilde{\cal H}_{\mu\nu\rho\,mn} 
\;.
\label{defdC}
\eea
It is a useful consistency test of the present construction, that with the reduction ansatz
described in the previous sections, the r.h.s.\ of this equation indeed takes the form of
a curl in the internal variables. Let us verify this explicitly.
Since the reduction ansatz is covariant, the first term reduces according to the form
of its free indices $[mn]$, c.f.~(\ref{BB2})
\bea
e\,
{\cal M}_{mn,N}\, {\cal F}^{\sigma\tau\,N}
&=&
-\frac12\,\sqrt{2}\, \partial_{[m} {\cal K}^{[ab]}{}_{n]}\left(
\sqrt{|{\rm g}|} \,M_{ab,N}\,F^{\sigma\tau\,N}\right)
\;,
\eea
which indeed takes the form of a curl.
We recall that the $D=5$ field strength $F_{\mu\nu}{}^{N}$ combines the 15
non-abelian field strengths $F_{\mu\nu}{}^{ab}$ and the 12 two-forms $B_{\mu\nu\,a\alpha}$
according to (\ref{F12}).
The reduction of the second term on the r.h.s.\ of (\ref{defdC}) is less obvious,
since $\tilde{\cal H}_{\mu\nu\rho\,mn}$ is not a manifestly covariant object,
and we have computed it explicitly by combining its defining equation (\ref{Htilde})
with the reduction of the vector fields (\ref{AA1}) and field strengths (\ref{FKK}).
With the identity (\ref{idKZ1}) among the Killing vectors and tensors, the second term on the r.h.s.\ of (\ref{defdC})
then reduces according to
\bea
\tilde{\cal H}_{\mu\nu\rho\,mn} &=&
\frac18\,\varepsilon_{abcdef}\,{\cal K}^{[ef]}{}_{mn}\,
\Omega_{\mu\nu\rho}^{abcd}
+ 2\,\partial_{[m}\left( {\cal A}_{[\mu}{}^{k}  {\cal A}_\nu{}^l {\cal A}_{\rho]n]kl} \right)
\;.
\label{Htildered}
\eea
with the non-abelian ${\rm SO}(p,q)$ Chern-Simons form defined as
\bea
\Omega_{\mu\nu\rho}^{abcd}
&=& \partial_{[\mu} A_{\nu}{}^{ab} A_{\rho]}{}^{cd}
+ F_{[\mu\nu}{}^{ab} A_{\rho]}{}^{cd}
\;,
\label{CSform}
\eea
in terms of the ${\rm SO}(p,6-p)$ Yang-Mills field strength $F_{\mu\nu}{}^{ab}$.
Again, (\ref{Htildered}) takes the form of a curl in the internal variables, such that
equation (\ref{defdC}) can be explicitly integrated to
\bea
C_{m\,\mu\nu\rho} &=& 
-\frac1{32}\,{\cal K}^{[ab]}{}_{m} \left(
2\,\sqrt{|{\rm g}|} \,\varepsilon_{\mu\nu\rho\sigma\tau}\,M_{ab,N} F^{\sigma\tau\,N}
+\sqrt{2}\,\varepsilon_{abcdef}\,
\Omega_{\mu\nu\rho}^{cdef} \right) 
\nonumber\\
&&{}
- \frac14\sqrt{2}\,{\cal K}_{[ab]}{}^k {\cal K}_{[cd]}{}^l {\cal Z}_{[ef]\,mkl}\left(
A_{[\mu}{}^{ab}  A_{\vphantom{[}\nu}{}^{cd} A_{\rho]}{}^{ef} \right)
\;.
\label{defC}
\eea
This yields the full reduction ansatz for the component $C_{m\,\mu\nu\rho}$.
Obviously, $C_{m\,\mu\nu\rho}$ is determined by (\ref{defdC}) only up to a gradient
$\partial_m \Lambda_{\mu\nu\rho}$ in the internal variables,
which corresponds to a gauge transformation of the IIB four-form.
Choosing the reduction ansatz (\ref{defC}), we have thus made a particular
choice for this gauge freedom.

In a similar way, the last missing component $C_{\mu\nu\rho\sigma}$ can be
reconstructed by further manipulating the equations and comparing to the IIB
self-duality equations~\cite{Baguet:2015xha}. Concretely, taking the external curl of (\ref{defdC}) and using
Bianchi identities and field equations on the r.h.s.\ yields a differential equation that can be
integrated in the internal variable to
\bea
-\frac1{6} \,e\varepsilon_{\mu\nu\rho\sigma\lambda}\,\varepsilon^{kpqrs}\,({\rm det}\,G)^{-1}\,G_{nk}\,\widehat{\cal D}^\lambda C_{pqrs} 
&=&
16\, D^{\KK}_{[\mu}  C_{\nu\rho\sigma]\,n} 
-30\,
 \varepsilon_{\alpha\beta}\,{\cal B}_{[\mu\nu}{}^\alpha\partial_n {\cal B}_{\rho\sigma]}{}^\beta 
\nonumber\\
&&{}
+6\,\sqrt{2}\,{\cal F}_{[\mu\nu}{}^k   {\cal A}_{\rho}{}^l {\cal A}_{\sigma]\,lkn}
+4\,\partial_n C_{\mu\nu\rho\sigma}
\;,\quad
\label{dualityD}
\eea
up to an external gradient $\partial_n C_{\mu\nu\rho\sigma}$ which carries the last
missing component of the IIB four-form.
Here, $D^\KK_\mu$ denotes the Kaluza-Klein covariant derivative 
\bea
D^\KK_\mu C_n &\equiv&
\partial_\mu C_n - {\cal A}_\mu{}^k \partial_k C_n - \partial_n {\cal A}_\mu{}^k C_k
\;,\qquad{\rm etc.}\;,
\label{DKK}
\eea
and
$\widehat{\cal D}_\mu C_{pqrs}$ is a particular combination of scalar covariant 
derivatives~\cite{Baguet:2015xha}, which is most compactly defined via particular
components of the scalar currents as
\bea
{\cal D}_\mu {\cal M}_{mn,N} {\cal M}^{N\,n}
&=&
\frac{\sqrt{2}}{3}\,({\rm det}\,G)^{-1}\,G_{mn}\, 
\varepsilon^{npqrs}\,\widehat{\cal D}_{\mu} C_{pqrs}
\;,
\label{DMMDc}
\eea
where ${\cal D}_\mu$ refers to the full EFT derivative, covariant under generalized diffeomorphisms. 
Again, it is a useful consistency check of the construction that with the reduction ansatz developed
so far, equation (\ref{dualityD}) indeed turns into a total gradient, from which we may read off 
the function $C_{\mu\nu\rho\sigma}$\,.
For the l.h.s.\ this is most conveniently seen by virtue of (\ref{DMMDc}) and the reduction ansatz
(\ref{SSembedding}) for ${\cal M}_{MN}$, giving rise to 
\bea
-4\,e \, ({\rm det}\,G)^{-1}\,G_{mk}\, \varepsilon^{kpqrs}\,\widehat{\cal D}^{\mu} C_{pqrs}
&=&
3\,\sqrt{|{\rm g}|} \,{\cal K}^{[ab]}{}_{mn}{\cal K}_{[cd]}{}^n
{D}^\mu {M}_{ab,N} {M}^{N\,cd}
\nonumber\\
&=&
6\,\sqrt{|{\rm g}|} \left(\sqrt{2}\,  {\cal K}^{[cb]}{}_{m} \eta_{ac} 
- \partial_m ( {\cal Y}^{b} {\cal Y}_{a} )  \right)
{D}^\mu {M}_{bd,N} {M}^{N\,da}
\;,\qquad
\eea
where we have used (\ref{idKK2}). The derivatives $D_\mu$ on the r.h.s.\ now refer
to the ${\rm SO}(p,6-p)$ covariant derivatives (\ref{covSO}).
For the terms on the r.h.s.\ of (\ref{dualityD}), we find
with (\ref{AA1}), (\ref{BB1}), and (\ref{idKYY})
\bea
-30\,
 \varepsilon_{\alpha\beta}\,{\cal B}_{[\mu\nu}{}^\alpha\partial_n {\cal B}_{\rho\sigma]}{}^\beta 
&=&
15\,\sqrt{2}\,
 \varepsilon_{\alpha\beta}\,{B}_{[\mu\nu}{}^{a\alpha} {B}_{\rho\sigma]}{}^{b\beta}\,
 {\cal K}_{[ab]\,n} 
 \;,
\nonumber\\
6\,\sqrt{2}\,{\cal F}_{[\mu\nu}{}^k   {\cal A}_{\rho}{}^l {\cal A}_{\sigma]\,lkn}
&=&
-6\sqrt{2}\,
{F}_{[\mu\nu}{}^{ab} 
A_\rho{}^{cd} A_{\sigma]}{}^{ef}\,
{\cal K}_{[ab]}{}^k   {\cal K}_{[cd]}{}^l {\cal Z}_{[ef]\,nkl}
\;,
\label{BBFF}
\eea
as well as
\bea
16\,D^\KK_{[\mu}C_{\nu\rho\sigma]\,m} &=&
\frac1{{2}}\,{\cal K}^{[ab]}{}_{m}\left(
\sqrt{|{\rm g}|}\,\varepsilon_{\mu\nu\rho\sigma\tau}\,
D_{\lambda} \left({M}_{ab,N} {F}^{\tau\lambda\,N}\right)
+\sqrt{2}\,\varepsilon_{abcdef}\,D_{[\mu} \Omega_{\nu\rho\sigma]}^{cdef} \right)
\nonumber\\
&&{}
+4\sqrt{2}\, {\cal F}_{[\mu\nu}{}^k  {\cal A}_\rho{}^l {\cal A}_{\sigma]\,mkl} 
+2\sqrt{2}\, {\cal A}_{[\mu}{}^k  {\cal A}_\nu{}^l {\cal F}_{\rho\sigma]\,mkl} 
\nonumber\\
&&{}
+\sqrt{2}\, {\cal A}_{[\mu}{}^k  {\cal A}_\nu{}^l
\left(
2\,{\cal A}_{\rho}{}^n \partial_{|n|}  {\cal A}_{\sigma]\,klm}
+3\,  \partial_{[m} {\cal A}_{\rho}{}^n {\cal A}_{\sigma]\,kl]n}
-3\, {\cal A}_{\rho}{}^n \partial_{[m}  {\cal A}_{\sigma]\,kl]n}
\right)
\nonumber\\
&&{}
-2\sqrt{2}\,{\cal A}_{[\mu}{}^k {\cal A}_{\nu\,|kmn|}  \left( {\cal A}_\rho{}^l  \partial_{|l|} {\cal A}_{\sigma]}{}^n \right)
-\sqrt{2}\, \partial_m \left({\cal A}_{[\mu}{}^k  {\cal A}_\nu{}^l  {\cal A}_{\rho}{}^n {\cal A}_{\sigma]\,kln}\right)
\;,
\label{DC4}
\eea
where we have explicitly evaluated the Kaluza-Klein covariant derivative $D_\mu$ on $C_{\mu\nu\rho\,m}$,
the latter given by (\ref{defC}). Moreover, we have arranged the ${\cal A}^4$ terms such that they allow for a 
convenient evaluation of their reduction formulae.
Namely, in the last two lines we have factored out the quadratic polynomials that correspond to the ${\cal A}^2$ terms
in the non-abelian field strengths (\ref{FKK}) and thus upon reduction factor in analogy to the field strengths,
leaving us with the ${A}^4$ terms
\bea
{\cal A}{\cal A}{\cal A}{\cal A}&\longrightarrow&
-2\, f_{ef,gh}{}^{ij} \, {\cal K}_{[ab]}^k
\left( {\cal Z}_{[cd]\,mkl}  {\cal K}_{[ij]}^l +  {\cal K}_{[cd]}^l {\cal Z}_{[ij]\,mkl} \right)
A_{[\mu}{}^{ab} A_{\nu}{}^{cd} A_\rho{}^{ef} A_{\sigma]}{}^{gh}
\nonumber\\
&&{}
-\sqrt{2}\,  \partial_m \left({\cal A}_{[\mu}{}^k  {\cal A}_\nu{}^l  {\cal A}_{\rho}{}^n {\cal A}_{\sigma]\,kln}\right)
\nonumber\\
&=&
-\frac14\,\sqrt{2}\, f_{ab,uv}{}^{xy} f_{ef,gh}{}^{ij} \,\varepsilon_{cdijxy}\,{\cal K}^{[uv]}_m
A_{[\mu}{}^{ab} A_{\nu}{}^{cd} A_\rho{}^{ef} A_{\sigma]}{}^{gh}
\nonumber\\
&&{}
+\frac12\,f_{ef,gh}{}^{ij} \,\varepsilon_{cdijau}\, \partial_m\left( {\cal Y}^u{\cal Y}_b\right) 
A_{[\mu}{}^{ab} A_{\nu}{}^{cd} A_\rho{}^{ef} A_{\sigma]}{}^{gh}
\nonumber\\
&&{}
-\sqrt{2}\,  \partial_m \left({\cal A}_{[\mu}{}^k  {\cal A}_\nu{}^l  {\cal A}_{\rho}{}^n {\cal A}_{\sigma]\,kln}\right)
\;,
\eea
upon using the identities (\ref{idKZ1}), (\ref{idKK2}).
While the last two terms are total gradients, the first term cancels against the corresponding contribution
from the derivative of the Chern-Simons form $\Omega^{abcd}_{\mu\nu\rho}$ in (\ref{DC4})
\bea
D^{\vphantom{a}}_{[\mu} \Omega^{cdef}_{\nu\rho\sigma]}\,\varepsilon_{abcdef}
&=&
\frac34\,{ F}_{[\mu\nu}{}^{cd} { F}_{\rho\sigma]}{}^{ef}\,\varepsilon_{abcdef}
-\frac12\,\sqrt{2}\,A_{[\mu}{}^{cd}A_{\nu\vphantom{]}}{}^{ef} {F}_{\rho\sigma]}{}^{gh}\,
f_{ab,ef}{}^{uv}\,\varepsilon_{cdghuv}
\nonumber\\
&&{}
-\frac12\,A_{[\mu}{}^{cd}
A_{\nu\vphantom{]}}{}^{ef}
A_{\rho\vphantom{]}}{}^{gh}
A_{\sigma]}{}^{ij}\,
f_{cd,ef}{}^{rs}
f_{gh,ij}{}^{uv}
\varepsilon_{abrsuv}
\;.
\label{DOmega}
\eea
Similarly, the ${\cal FAA}$ terms in (\ref{DC4}) combine with those of (\ref{BBFF}) according to
\bea
{\cal FAA} &\longrightarrow&
-2\sqrt{2}\,
{F}_{[\mu\nu}{}^{ab} A_\rho{}^{cd} A_{\sigma]}{}^{ef}\, {\cal K}_{[cd]}^l
\left(
{\cal K}_{[ab]}^k   {\cal Z}_{[ef]\,mkl}+{\cal K}_{[ef]}^k {\cal Z}_{[ab]\,mkl}
\right)
\\
&=&
\frac12\,f_{cd,ij}{}^{gh}\,{\cal K}^{[ij]}{}_m
{F}_{[\mu\nu}{}^{ab} A_\rho{}^{cd} A_{\sigma]}{}^{ef}\, 
\varepsilon_{abefgh}\,
-\frac12\,\sqrt{2}\,
{F}_{[\mu\nu}{}^{ab} A_\rho{}^{cd} A_{\sigma]}{}^{ef}\, 
\varepsilon_{abefch}\, \partial_m({\cal Y}^h {\cal Y}_d)
\;.
\nonumber
\eea
Again, the first term cancels against the corresponding contribution
from the derivative of the Chern-Simons form $\Omega^{abcd}_{\mu\nu\rho}$, given in (\ref{DOmega}).

Collecting all the remaining terms, 
equation (\ref{dualityD}) takes the final form
\bea
0&=&
\frac12\, {\cal K}^{[ab]}{}_{m} \,
\sqrt{|{\rm g}|}\,\varepsilon_{\mu\nu\rho\sigma\tau}
\left(
\frac12\,\sqrt{2}\,\eta_{da} 
\,{D}^\tau {M}_{cb,N} {M}^{N\,cd} + D_{\lambda} \left({M}_{ab,N} {F}^{\tau\lambda\,N}\right)
\right)
\nonumber\\
&&{}+\frac38\, \sqrt{2}\,
{\cal K}^{[ab]}{}_{m}\left(
\varepsilon_{abcdef}\,{ F}_{[\mu\nu}{}^{cd} { F}_{\rho\sigma]}{}^{ef} 
+40\,
 \varepsilon_{\alpha\beta}\,\eta_{ac}\eta_{bd}\,{B}_{[\mu\nu}{}^{c\alpha} {B}_{\rho\sigma]}{}^{d\beta}
\right)
\nonumber\\
&&{}
+\frac12\,f_{ef,gh}{}^{ij} \,\varepsilon_{cdijay}\, \partial_m\left( {\cal Y}^y{\cal Y}_b\right) 
A_{[\mu}{}^{ab} A_{\nu}{}^{cd} A_\rho{}^{ef} A_{\sigma]}{}^{gh}
\nonumber\\
&&{}
-\frac14\,\sqrt{|{\rm g}|}\,\varepsilon_{\mu\nu\rho\sigma\tau} \, \partial_m ( {\cal Y}^{b} {\cal Y}_{d} )\,
{D}^\tau {M}_{ab,N} {M}^{N\,ad}
-\frac12\,\sqrt{2}\,{F}_{[\mu\nu}{}^{ab} A_\rho{}^{cd} A_{\sigma]}{}^{ef}\, 
\varepsilon_{abefch}\, \partial_m({\cal Y}^h {\cal Y}_d)
\nonumber\\
&&{}
-\sqrt{2}\, \partial_m \left({\cal A}_{[\mu}{}^k  {\cal A}_\nu{}^l  {\cal A}_{\rho}{}^n {\cal A}_{\sigma]\,kln}\right)
+4\,\partial_m C_{\mu\nu\rho\sigma}
\;.
\label{DC5}
\eea
Now the first two lines of the expression
precisely correspond to the vector field equations 
(\ref{eomV}) of the $D=5$ theory,
which confirms that on-shell this equation reduces
to a total gradient in the internal variables.
Although guaranteed by the consistency of the generalized Scherk-Schwarz ansatz
and the general analysis of~\cite{Baguet:2015xha}, it is gratifying that this structure is confirmed 
by explicit calculation based on the $D=5$ field equations and
the non-trivial identities among the Killing vectors.
We are thus in position to read off from (\ref{DC5}) the final expression for the 4-form as
\bea
C_{\mu\nu\rho\sigma} &=&
-\frac1{16}\, {\cal Y}_{a}  {\cal Y}^{b}
 \left(
\sqrt{|{\rm g}|}\,\varepsilon_{\mu\nu\rho\sigma\tau}
{D}^\tau {M}_{bc,N} {M}^{N\,ca}\, 
+2\,\sqrt{2}\,\varepsilon_{cdefgb}\, {F}_{[\mu\nu}{}^{cd} A_\rho{}^{ef} A_{\sigma]}{}^{ga}\, 
\right)
\nonumber\\
&&{}
+\frac14\,\left(\sqrt{2}\, {\cal K}_{[ab]}{}^k {\cal K}_{[cd]}{}^l {\cal K}_{[ef]}{}^n {\cal Z}_{[gh]\,kln}
-{\cal Y}_{h}  {\cal Y}^{j}\,\varepsilon_{abcegj}\, \eta_{df}\right)
A_{[\mu}{}^{ab} A_{\vphantom{]}\nu}{}^{cd} A_{\vphantom{]}\rho}{}^{ef} A_{\sigma]}{}^{gh}
\nonumber\\
&&{}
+ \Lambda_{\mu\nu\rho\sigma}(x)
\;,
\label{C4}
\eea
in terms of the $D=5$ fields,
up to an $y$-independent term $\Lambda_{\mu\nu\rho\sigma}(x)$,
left undetermined by equation (\ref{dualityD}) and 
fixed by the last component of the IIB self-duality equations~(\ref{FF55}).
This equation translates into
\bea
4\,D^\KK_{[\mu} C^{}_{\nu\rho\sigma\tau]}
&=&
30\,
 \varepsilon_{\alpha\beta}\,{\cal B}_{[\mu\nu}{}^\alpha D^\KK_\rho {\cal B}_{\sigma\tau]}{}^\beta 
+8\,   {\cal F}_{[\mu\nu}{}^k   C_{\rho\sigma\tau]\,k} \nonumber\\
&&-
\frac{1}{120}\, e\varepsilon_{\mu\nu\rho\sigma\tau}\varepsilon^{klmnp}\,({\rm det}\,G)^{-4/3} X_{klmnp}
\;,
\label{FF5}
\eea
where $X_{klmnp}$ is a combination of internal derivatives of the scalar fields, c.f.~\cite{Baguet:2015xha},
that is most compactly given by
\bea
\frac{1}{120}\,
 \varepsilon^{kpqrs}\,X_{kpqrs}
 &=&
 -\frac1{20}\,\sqrt{2}\,({\rm det}\,G) \, G^{ml} \, \partial_l  {\cal M}_{mn,N} {\cal M}^{N\,n}
\;,
\label{XXMM}
\eea
in analogy to (\ref{DMMDc}).
It can be shown that equation (\ref{FF5}) can be derived from the external curl of equations (\ref{dualityD})
upon using the EFT field equations and Bianchi identities, up to a $y$-independent equation that defines
the last missing function $\Lambda_{\mu\nu\rho\sigma}$. For the general case this has been worked out 
in~\cite{Baguet:2015xha}. Alternatively, it can be confirmed by explicit calculation with the Scherk-Schwarz reduction ansatz,
that equation (\ref{FF5}) with the components $C_{\mu\nu\rho\,m}$ and $C_{\mu\nu\rho\sigma}$
from (\ref{defC}) and (\ref{C4}), respectively, decomposes into a $y$-dependent part, which vanishes due to
the $D=5$ scalar equations of motion, and a $y$-independent part, that defines the function $\Lambda_{\mu\nu\rho\sigma}$.
The calculation is similar (but more lengthy) than the previous steps, requires the same non-trivial identities
among Killing vectors derived above, but also some non-trivial algebraic identities among the components of the
scalar ${\rm E}_{6(6)}$ matrix ${M}_{MN}$. We relegate the rather lengthy details to appendix~\ref{app:lambda} and simply report
the final result from equation (\ref{LambdaA})
\bea
D_{[\mu} \Lambda_{\nu\rho\sigma\tau]}
&=& 
-\frac1{480}\, 
\,\sqrt{|{\rm g}|}\,\varepsilon_{\mu\nu\rho\sigma\tau} D_{\lambda}
\left({M}^{N\,ac} \,{D}^\lambda {M}_{ac,N} \right) 
\nonumber\\
&&{} 
+\frac{1}{240}\, 
\sqrt{|{\rm g}|}\,\varepsilon_{\mu\nu\rho\sigma\tau}\,{F}^{\kappa\lambda\,N}\left(
{M}_{ab,N}  {F}_{\kappa\lambda}{}^{ab}
-\frac12\,\sqrt{10}\,\varepsilon_{\alpha\beta}\,\eta_{ab}\,M^{a\,\alpha}{}_N\,
{B}_{\kappa\lambda}{}^{b\,\beta}  
\right)
\nonumber\\
&&{}
+\frac1{600}\,\sqrt{|{\rm g}|}\,\varepsilon_{\mu\nu\rho\sigma\tau}
 \left(
10\, \delta_h^d\delta^a_e
+   2\,{M}^{fd,ga}  {M}_{gh,fe}  -
{M}_{e\alpha}{}^{ga} 
 {M}_{gh}{}^{d\alpha}
\right) M^{bh,ec}\,\eta_{cd} \eta_{ab}
\nonumber\\
&&{}
+\frac1{32}\,\sqrt{2}\, 
 \,\varepsilon_{abcdef}\,  {F}_{[\mu\nu}{}^{ab} F_{\rho\sigma}{}^{cd}  A_{\tau]}{}^{ef}
+ \frac1{16}\,
 {F}_{[\mu\nu}{}^{ab} A_\rho{}^{cd} A_{\sigma}{}^{ef}  A_{\tau]}{}^{gh} \,\varepsilon_{abcdeh} \eta_{fh}
 \nonumber\\
&&{}
+ \frac1{40}\,\sqrt{2}\,\, A_{[\mu}{}^{ab} A_{\nu}{}^{cd}A_\rho{}^{ef} A_\sigma{}^{gh} A_{\tau]}{}^{ij}\,
 \varepsilon_{abcegi}\,\eta_{df}\eta_{hj}
\;.
\label{defLambda}
\eea
Since there is no non-trivial Bianchi identity for (\ref{defLambda}), this equation can be integrated
and yields the last missing term in the four-form potential (\ref{C4}). This completes
the reduction formulae for the full set of fundamental IIB fields.

\section{Summary}

We have in this paper derived the explicit reduction formulae
for the full set of IIB fields in the compactification on the sphere $S^5$ and
the inhomogeneous hyperboloids $H^{p,6-p}$. The fluctuations around the background geometry
are described by a $D=5$ maximal supergravity, with gauge group ${\rm SO}(p,6-p)$\,.
The dependence on the internal variables is explicitly expressed in terms of 1) a set
of vectors ${\cal K}_{[ab]}{}^m$ which are Killing vectors of a homogeneous 
metric $\tilde{G}_{mn}$ (\ref{Gtilde}), and 2) a four-form $\tilde{C}_{\mu\nu\rho\sigma}$ 
whose field-strength yields the Lorentzian volume form (\ref{dettildec}).
Only for the compact case of $S^5$, the metric $\tilde{G}_{mn}$ and four-form $\tilde{C}_{klmn}$
coincide with the space-time background geometry.
In the non-compact case, they refer to a (virtual) homogeneous Lorentzian geometry which encodes 
the inhomogeneous space-time background geometry via the formulas provided. This is in accordance with the ansatz
proposed and tested for some stationary points of the non-compact $D=4$ gaugings in \cite{Baron:2014bya},
see also \cite{Hull:1988jw,Cvetic:2004km} for earlier work.
Only for $p=6$ and $p=3$ does the background geometry provide a solution to the IIB field equations.
We stress, that also in the remaining cases, the reduction ansatz describes a consistent truncation of the
IIB theory to an effectively $D=5$ supergravity theory, just this theory does not have a simple ground state 
with all fields vanishing. Still, any stationary point or holographic RG flow 
of these non-compact gaugings as well as any other solution to their field equations lifts to a IIB solution
by virtue of the explicit reduction formulas.

The explicit reduction formulas are derived via the EFT formulation of the IIB theory
by evaluating the formulas of the generalized Scherk-Schwarz reduction ansatz for
the twist matrices obtained in~\cite{Hohm:2014qga}. The Scherk-Schwarz origin also proves consistency of
the truncation in the sense that all solutions of the respective $D=5$ maximal supergravities
lift to solutions of the type IIB fields equations. 
By virtue of the explicit embedding of the IIB theory into EFT \cite{Hohm:2013vpa,Baguet:2015xha} 
these formulas can be pulled back to read off the reduction formulas for the original type IIB fields.
Upon some further computational effort we have also derived the explicit expressions for all the components
of the IIB four-form. Along the way, we explicitly verified the IIB self-duality equations. Although their
consistency is guaranteed by the general construction, we have seen that their validation by virtue of non-trivial
Killing vector identities still represents a rewarding exercise.

We have in this paper restricted the construction to the bosonic sector of type IIB supergravity.
In the EFT framework, consistency of the reduction of the fermionic sector follows along the same lines 
from the supersymmetric extension of the E$_{6(6)}$ exceptional field theory \cite{Musaev:2014lna} which upon generalized
Scherk-Schwarz reduction yields the fermionic sector of the $D=5$ gauged supergravities~\cite{Hohm:2014qga}.
In particular, compared to the bosonic reduction ansatz~(\ref{SSembedding}), the EFT fermions reduce as
scalar densities, i.e.\ their $y$-dependence is carried by some power of the scale factor,
such as $\psi_{\mu}{}^{i}(x,y) \ = \ \rho^{-\frac{1}{2}}(y)\,\psi_{\mu}{}^{i}(x) $, etc..
A derivation of the explicit reduction formulas for the original IIB fermions would require the
dictionary of the fermionic sector of EFT into the IIB theory, presumably along the lines of \cite{Ciceri:2014wya}.
The very existence of a consistent reduction of the fermionic sector can also be inferred on general grounds~\cite{Cvetic:2000dm}
combining the bosonic results with the supersymmetry of the IIB theory.

We close by recollecting the full set of IIB reduction formulas derived in this paper.
The IIB metric is given by
\bea
ds^2 &=&
 \Delta^{-2/3}(x,y)\,{\rm g}_{\mu\nu}(x) \, dx^\mu dx^\nu  
\nonumber\\
 &&{}
 + 
G_{mn}(x,y)
\left( dy^m + {\cal K}_{[ab]}{}^m(y) A_\mu^{ab}(x)  dx^\mu\right)\left( dy^n +{\cal K}_{[cd]}{}^n(y) A_\nu^{cd}(x)   dx^\nu\right)
 \;,
 \label{metricIIBsumm}
\eea
in standard Kaluza-Klein form, in terms of vectors ${\cal K}_{[ab]}{}^m$ from (\ref{KV}) that
are Killing for the (Lorentzian) metric $\tilde{G}_{mn}$ from (\ref{Gtilde}), and
the internal block $G_{mn}$ of the metric (\ref{metricIIBsumm}) given by the inverse of 
\bea
 G^{mn}(x,y) &=& \,\Delta^{2/3}(x,y)\, {\cal K}_{[ab]}{}^m(y) {\cal K}_{[cd]}{}^n(y)\,M^{ab,cd}(x)
\;.
\eea
The IIB dilaton and axion combine into the symmetric ${\rm SL}(2)$ matrix
\bea
m^{\alpha\beta}(x,y) &=&  
 \Delta^{4/3}(x,y)\, {\cal Y}_a(y) {\cal Y}_b(y) \, M^{a\alpha,b\beta}(x)
\;,
\label{DAsumm}
\eea
in terms of the harmonics ${\cal Y}_a$ from (\ref{defY}).
Since ${\rm det}\,m^{\alpha\beta}=1$, this equation can also be used as
a defining equation for the function $\Delta(x,y)$\,.
The different components of the two-form doublet are given by
\bea
C_{mn}{}^\alpha(x,y)
 &=&
-\frac12\,  
\varepsilon^{\alpha\beta} \Delta^{4/3}(x,y)\, m_{\beta\gamma}(x,y)\, {\cal Y}_c(y)\,{\cal K}^{[ab]}{}_{mn}(y)
\,{M}_{ab}{}^{c\gamma}(x)
\;,
\nonumber\\
{C}_{\mu m}{}^{\alpha}(x,y) &=&0\;,
\nonumber\\
{C}_{\mu \nu}{}^{\alpha}(x,y) &=&\sqrt{10}\, {\cal Y}_a(y)\,B_{\mu\nu}{}^{a\alpha}(x)
\;.
\label{C2summ}
\eea
Next, we give the uplift formulas for the four-form components
in terms of the Killing vectors ${\cal K}_{[ab]}{}^m(y)$, Killing tensors ${\cal K}_{[ab]\,mn}(y)$, the 
sphere harmonics ${\cal Y}_a(y)$ given in (\ref{defY}),
the function ${\cal Z}_{[ab]\,kmn}(y)$ given by (\ref{defZ}), and the four-form $\tilde{C}_{klmn}(y)$ from (\ref{Ctilde}).
In order not to clutter the formulas, in the following 
we do not display the dependence on the arguments $x$ and $y$ as it is always clear from the definition 
 of the various objects whether they depend on the external or internal coordinates or both. 
The final result reads 
\bea
C_{klmn} &=&  \tilde{C}_{klmn} +
\frac{1}{16}\,\tilde\omega_{klmnp}\,   \Delta^{4/3}\,m_{\alpha\beta} \,
 \tilde{G}^{pq} \, \partial_q \left(\Delta^{-4/3} \,m^{\alpha\beta}\right) 
 \;,
 \nonumber\\
C_{\mu kmn}&=&\frac{\sqrt{2}}{4}\,\mathcal{Z}_{[ab] \,kmn}\,A_\mu{}^{ab}
\;,\nonumber\\
C_{\mu \nu \,mn}&=&\frac{\sqrt{2}}{4}\,
{\cal K}_{[ab]}{}^k {\cal Z}_{[cd]\,kmn}\,
{A}_{[\mu}{}^{ab} {A}_{\nu]}{}^{cd}\;,
\quad
\nonumber\\
 C_{m\,\mu\nu\rho} &=& 
-\frac1{32}\,{\cal K}_{[ab]\,m}\left(
2\,\sqrt{|{\rm g}|} \,\varepsilon_{\mu\nu\rho\sigma\tau}\,M_{ab,N} F^{\sigma\tau\,N}
+\sqrt{2}\,\varepsilon_{abcdef}\,
\Omega_{\mu\nu\rho}^{cdef} \right) 
\nonumber\\
&&{}
- \frac14\sqrt{2}\,{\cal K}_{[ab]}{}^k {\cal K}_{[cd]}{}^l{\cal Z}_{[ef]\,mkl}\left(
A_{[\mu}{}^{ab}  A_{\vphantom{[}\nu}{}^{cd} A_{\rho]}{}^{ef} \right)
\;,\nonumber\\
 C_{\mu\nu\rho\sigma} &=&
-\frac1{16}\, {\cal Y}_{a}  {\cal Y}^{b}
 \left(
\sqrt{|{\rm g}|}\,\varepsilon_{\mu\nu\rho\sigma\tau}
{D}^\tau {M}_{bc,N} {M}^{N\,ca}\, 
+2\,\sqrt{2}\,\varepsilon_{cdefgb}\, {F}_{[\mu\nu}{}^{cd} A_{\vphantom{]}\rho}{}^{ef} A_{\vphantom{]}\sigma]}{}^{ga}\, 
\right)
\nonumber\\
&&{}
+\frac14\,\left(\sqrt{2}\, {\cal K}_{[ab]}{}^k {\cal K}_{[cd]}{}^l {\cal K}_{[ef]}{}^n{\cal Z}_{[gh]\,kln}
-{\cal Y}_{h}  {\cal Y}^{j}\,\varepsilon_{abcegj}\, \eta_{df}\right)
A_{[\mu}{}^{ab} A_{\vphantom{]}\nu}{}^{cd} A_{\vphantom{]}\rho}{}^{ef} A_{\sigma]}{}^{gh}
\nonumber\\
&&{}
+ \Lambda_{\mu\nu\rho\sigma}(x)
\;. 
 \label{C4summ}
\eea
We recall, that the curved indices on these objects are raised and lowered with the $x$-independent
metric $\tilde{G}_{mn}(y)$ from (\ref{Gtilde})
and not with the background metric $G_{mn}$.
The function $\Lambda_{\mu\nu\rho\sigma}(x)$ is defined by equation (\ref{defLambda}).
All $p$-form components are given in the basis after standard Kaluza-Klein decomposition, 
explicitly related to the original IIB fields by (\ref{C2C4}).

With the reduction ansatz (\ref{metricIIBsumm})--(\ref{C4summ}), the type IIB field equations reduce
to the $D=5$ field equations derived from the Lagrangian (\ref{LD5}). As a consequence,
these formulas lift every solution
of $D=5$, ${\rm SO}(p,q)$ gauged supergravity to a solution of IIB supergravity.

\subsection*{Acknowledgements}
We would like to thank Bernard de Wit, Edvard Musaev, and
Mario Trigiante  for helpful discussions.
The work of O.H.~is supported by the U.S. Department of Energy (DoE) under 
the cooperative research agreement DE-FG02-05ER41360 and a DFG Heisenberg fellowship.

\section*{Appendix}

\begin{appendix}

\section{Finding $\Lambda_{\mu\nu\rho\sigma}$}
\label{app:lambda}

In order to find the last missing contribution $\Lambda_{\mu\nu\rho\sigma}$ in the expression (\ref{C4}) for the four-form
component $C_{\mu\nu\rho\sigma}$ let us study the reduction of the different terms of equation (\ref{FF5})
\bea
\frac{1}{120}\, e\varepsilon_{\mu\nu\rho\sigma\tau}\varepsilon^{klmnp}\,({\rm det}\,G)^{-4/3} X_{klmnp}
&=&
30\,
 \varepsilon_{\alpha\beta}\,{\cal B}_{[\mu\nu}{}^\alpha D^\KK_{\rho} {\cal B}_{\sigma\tau]}{}^\beta 
+8\,   {\cal F}_{[\mu\nu}{}^k   C_{\rho\sigma\tau]\,k} 
\nonumber\\
&&{}
-4\,D^\KK_{[\mu} C_{\nu\rho\sigma\tau]}
\;. 
\label{FF5A}
\eea
By construction, after imposing the generalized Scherk-Schwarz ansatz this equation should split into
a $y$-dependent part proportional to the $D=5$ scalar field equations (\ref{D5scalarEOM}),  and a $y$-independent part
which determines the function $\Lambda_{\mu\nu\rho\sigma}$.

The first term on the r.h.s.\ simply reduces according to the reduction ansatz (\ref{BB1})
\bea
30\,
 \varepsilon_{\alpha\beta}\,{\cal B}_{[\mu\nu}{}^\alpha { D}^{{\vphantom{[}}\KK}_{\rho} {\cal B}_{\sigma\tau]}{}^\beta 
&=& 30\,\varepsilon_{\alpha\beta}\,\mathcal{Y}_a\mathcal{Y}_b\,{B}_{[\mu\nu}{}^{a\alpha} D_{\vphantom{[}\rho} {B}_{\sigma\tau]}{}^{b\beta}\;.
\eea
Note that the Kaluza-Klein covariant derivative turns into the ${\rm SO}(p,6-p)$ covariant
derivative by virtue of (\ref{idKY}).
With (\ref{defC}) and the identity \eqref{idKK}, we find for the second term on the r.h.s.\ of (\ref{FF5A})
\bea
8\,  {\cal F}_{[\mu\nu}{}^k   C_{\rho\sigma\tau]\,k} &=&
 -\frac12\,{\cal Y}_b {\cal Y}^a \, {F}_{[\mu \nu}{}^{cb}\left(
2\,\sqrt{|{\rm g}|}\,\varepsilon_{\rho\sigma\tau]\kappa\lambda}\,
{M}_{ac,N} {F}^{\kappa\lambda\,N}
+\sqrt{2}\,\Omega_{\rho\sigma\tau]}^{efgh}\,\varepsilon_{acefgh} \right)
\nonumber\\
&&{}
+2\,\sqrt{2}\, {F}_{[\mu \nu}{}^{ab} A_{\rho}{}^{cd}  A_\sigma{}^{ef} A_{\tau]}{}^{gh} \,
{\cal K}_{[ab]}{}^m{\cal K}_{[cd]}{}^k {\cal K}_{[ef]}{}^l {\cal Z}_{[gh]\,mkl}
\;.
\eea
Next, we have to work out the covariant curl of $C_{\mu\nu\rho\sigma}$ with the explicit expression (\ref{C4}).
To this end, we first note that for all terms with $y$-dependence proportional to ${\cal Y}^a {\cal Y}^b$, the Kaluza-Klein covariant
derivative reduces to
\bea
D_\mu^\KK \left({\cal Y}^a {\cal Y}^b\,X_{ab}\right) &=& {\cal Y}^a {\cal Y}^b\,D_\mu X_{ab}
\;,
\label{DKKD}
\eea
in view of the property (\ref{idKY}) of the harmonics ${\cal Y}^a$\,. We thus find
\bea
-4\,D_{[\mu}^\KK C_{\nu\rho\sigma\tau]} &=&
\frac1{20}\, {\cal Y}_{a}  {\cal Y}^{b}
\,\sqrt{|{\rm g}|}\,\varepsilon_{\mu\nu\rho\sigma\tau} D_{\lambda}
\left({M}^{N\,ca} \,{D}^\lambda {M}_{bc,N} \right) - 4\, D_{[\mu} \Lambda_{\nu\rho\sigma\tau]}
\nonumber\\
&&{}
+\frac1{2}\,\sqrt{2}\, {\cal Y}_{b}  {\cal Y}^{a}
 \,\varepsilon_{acdefg}\, D_{[\mu} \left( {F}_{\nu\rho}{}^{cd} A_\sigma{}^{ef} A_{\tau]}{}^{bg} 
 +\sqrt{2} \,
A_{\nu}{}^{cd} A_{\vphantom{]}\rho}{}^{eh} A_{\vphantom{]}\sigma}{}^{fj} A_{\tau]}{}^{bg}\, \eta_{hj} \right)
\nonumber\\
&&{}
-\sqrt{2}\,D_{[\mu}^\KK 
\left( {\cal K}_{[ab]}{}^k {\cal K}_{[cd]}{}^l {\cal K}_{[ef]}{}^n {\cal Z}_{[gh]\,kln}
A_{\nu}{}^{ab} A_{\vphantom{]}\rho}{}^{cd} A_{\vphantom{]}\sigma}{}^{ef} A_{\tau]}{}^{gh} \right)
\;.
\label{DC4A}
\eea
In order to evaluate the last term it is important to note that unlike in (\ref{DKKD}),
the Kaluza-Klein covariant derivative here cannot
just be pulled through the (non-covariant) $y$-dependent functions but has to be evaluated
explicitly leading to 
\bea
-\sqrt{2}\,D^\KK_{[\mu}\left({\cal A}_{\nu}{}^k {\cal A}_{\rho}{}^l {\cal A}_{\sigma}^n {\cal A}_{\tau]\, kln}\right)
&=&- \frac32\,\sqrt{2}\,
{F}_{[\mu\nu}{}^{ab}A_{\rho}{}^{cd}A_{\sigma}{}^{ef}A_{\tau]}{}^{gh} 
\, \mathcal{K}_{[ab]}{}^k \mathcal{K}_{[cd]}{}^l\mathcal{K}_{[ef]}{}^n \,\mathcal{Z}_{[gh] \, kln}\nonumber\\
&&{}+
\frac12\,\sqrt{2}\,
{F}_{[\mu\nu}{}^{ab}A_{\rho}{}^{cd}A_{\sigma}{}^{ef}A_{\tau]}{}^{gh} \,
\mathcal{K}_{[cd]}{}^k \mathcal{K}_{[ef]}{}^l \mathcal{K}_{[gh]}{}^n\,\mathcal{Z}_{[ab]\, kln}
\nonumber\\
&&{}
+\frac3{10}\,\sqrt{2}\,A_{[\mu}{}^{rs}A_{\nu}{}^{uv}A_{\rho}{}^{cd}A_{\sigma}{}^{ef}
A_{\tau]}{}^{gh}f_{cd,rs}{}^{ab}\varepsilon_{abuvge}\mathcal{Y}_f\mathcal{Y}_h
\;,
\nonumber
\eea
after some manipulation of the functions ${\cal K}_{[ab]}$, ${\cal Z}_{[ab]}$. 
Putting everything together and again using once more the identity (\ref{idKZ1}), the full r.h.s.\ of equation (\ref{FF5A})
is given by
\bea
\eqref{FF5A}_{{\rm r.h.s}} &=& 
\frac1{20}\, {\cal Y}_{a}  {\cal Y}^{b}
\,\sqrt{|{\rm g}|}\,\varepsilon_{\mu\nu\rho\sigma\tau} D_{\lambda}
\left({M}^{N\,ca} \,{D}^\lambda {M}_{bc,N} \right) - 4\, D_{[\mu} \Lambda_{\nu\rho\sigma\tau]}
\nonumber\\
&&{}
+\frac1{2}\,\sqrt{2}\, {\cal Y}_{a}  {\cal Y}^{b}
 \,\varepsilon_{bcdefg}\, D_{[\mu} \left( {F}_{\nu\rho}{}^{cd} A_\sigma{}^{ef} A_{\tau]}{}^{ag} 
 +\sqrt{2} \,
A_{\nu}{}^{cd} A_{\vphantom{]}\rho}{}^{eh} A_{\vphantom{]}\sigma}{}^{fj} A_{\tau]}{}^{ag}\, \eta_{hj} \right)
\nonumber\\
&&{}+\frac12\,\varepsilon_{dfghce}\mathcal{Y}_a\mathcal{Y}_b \,{ F}_{[\mu \nu}{}^{df} A_{\rho}{}^{ac}  A_\sigma{}^{be} A_{\tau]}{}^{gh}
+30\,
 \varepsilon_{\alpha\beta}\,\mathcal{Y}_a\mathcal{Y}_b\,{B}_{[\mu\nu}{}^{a\,\alpha} D_\rho {B}_{\sigma\tau]}{}^{b\,\beta}
\nonumber\\
&&{}
+\frac35\,\sqrt{2}\,\varepsilon_{csuvge}\,\mathcal{Y}_a\mathcal{Y}_b
\,\eta_{dr}\,A_{[\mu}{}^{rs}A_{\nu}{}^{uv}A_{\rho}{}^{cd}A_{\sigma}{}^{ae}A_{\tau]}{}^{bg}\,
\nonumber\\
&&{} -\frac12\,{\cal Y}_b {\cal Y}^a \, {F}_{[\mu \nu}{}^{cb}\left(
2\,\sqrt{|{\rm g}|}\,\varepsilon_{\rho\sigma\tau]\kappa\lambda}\,
{M}_{ac,N} {F}^{\kappa\lambda\,N}
+\sqrt{2}\,\Omega_{\rho\sigma\tau]}^{efgh}\,\varepsilon_{acefgh} \right) 
\;.
\eea
Some calculation and use of the Schouten identity shows that all terms carrying explicit gauge fields
add up precisely such that their $y$-dependence drops out due to ${\cal Y}_a {\cal Y}^a=1$\,.
Specifically, we find 
\bea
\eqref{FF5A}_{{\rm r.h.s}} \Big|_{FFA} &=& 
\frac1{8}\,\sqrt{2}\, 
 \,\varepsilon_{abcdef}\,  {F}_{[\mu\nu}{}^{ab} F_{\rho\sigma}{}^{cd}  A_{\tau]}{}^{ef}
 \;,\nonumber\\
 \eqref{FF5A}_{{\rm r.h.s}} \Big|_{FAAA} &=&  
 \frac14\,
 {F}_{[\mu\nu}{}^{ab} A_\rho{}^{cd} A_{\sigma}{}^{ef}  A_{\tau]}{}^{gh} \,\varepsilon_{abcdeh} \eta_{fh}
 \;,\nonumber\\
 \eqref{FF5A}_{{\rm r.h.s}} \Big|_{AAAAA} &=& 
\frac1{10}\,\sqrt{2}\, A_{[\mu}{}^{ab} A_{\nu}{}^{cd}A_\rho{}^{ef} A_\sigma{}^{gh} A_{\tau]}{}^{ij}\,
 \varepsilon_{abcegi}\,\eta_{df}\eta_{hj}
 \;.
\eea
In addition, we use the $D=5$ duality equation (\ref{duality5}) in order to rewrite the $BDB$ term of
(\ref{FF5A}) and arrive at
\bea
\eqref{FF5A}_{{\rm r.h.s}} &=& 
-\frac1{20}\, {\cal Y}_{a}  {\cal Y}^{b}
\,\sqrt{|{\rm g}|}\,\varepsilon_{\mu\nu\rho\sigma\tau} D_{\lambda}
\left({M}^{N\,ac} \,{D}^\lambda {M}_{bc,N} \right) - 4\, D_{[\mu} \Lambda_{\nu\rho\sigma\tau]}
\nonumber\\
&&{} _+
\frac{1}{10}\,
{\cal Y}_a {\cal Y}^b \, 
\sqrt{|{\rm g}|}\,\varepsilon_{\mu\nu\rho\sigma\tau}\,{F}^{\kappa\lambda\,N}\left(
{M}_{bc,N}  {F}_{\kappa\lambda}{}^{ac}
-\frac12\,\sqrt{10}\,\varepsilon_{\alpha\beta}\,\eta_{db}\,M^{d\,\alpha}{}_N\,
{B}_{\kappa\lambda}{}^{a\,\beta}  
\right)
\nonumber\\
&&{}
+\frac1{8}\,\sqrt{2}\, 
 \,\varepsilon_{abcdef}\,  {F}_{[\mu\nu}{}^{ab} F_{\rho\sigma}{}^{cd}  A_{\tau]}{}^{ef}
+ \frac14\,
 {F}_{[\mu\nu}{}^{ab} A_\rho{}^{cd} A_{\sigma}{}^{ef}  A_{\tau]}{}^{gh} \,\varepsilon_{abcdeh} \eta_{fh}
 \nonumber\\
&&{}
+\frac1{10}\,\sqrt{2}\, A_{[\mu}{}^{ab} A_{\nu}{}^{cd}A_\rho{}^{ef} A_\sigma{}^{gh} A_{\tau]}{}^{ij}\,
 \varepsilon_{abcegi}\,\eta_{df}\eta_{hj}
\;.\label{rhs_exp}
\eea
Structurewise, the r.h.s.\ of equation (\ref{FF5A}) is thus of the form
\bea
\eqref{FF5A}_{{\rm r.h.s}} &=& \left({\cal Y}_a(y) {\cal Y}_b(y) - \frac16\,\eta_{ab}\right){\cal E}_{1\,ab}(x)
+ {\cal E}_2(x)
\;.
\label{rhs}
\eea
Consistency of the reduction ansatz then implies that also the l.h.s.\ of (\ref{FF5A}) 
organizes into the same structure. The coefficients multiplying the $y$-dependent 
factor $\left({\cal Y}_a(y) {\cal Y}_b(y) - \frac16\,\eta_{ab}\right)$ must combine into a $D=5$ field equation
in order to reduce (\ref{FF5A}) to an $y$-independent equation which then 
provides the defining equation for $\Lambda_{\mu\nu\rho\sigma}$.

In order to see this explicitly, we recall, that the l.h.s.\ of (\ref{FF5A}) is defined by (\ref{XXMM}),
which together with the reduction ansatz (\ref{SSembedding}) for ${\cal M}_{MN}$ may be used to
read off the form of this term after reduction.
After some manipulation of the Killing vectors and tensors and use of the identities
collected in section~\ref{subsec:useful}, we obtain
\bea
\frac1{120}\,e\,\varepsilon^{klmnp}\,({\rm det}\,G)^{-4/3} X_{klmnp}
&=&
-\frac1{10}\,\sqrt{2} \,\sqrt{|{\rm g}|}\,{\cal Y}_{a} {\cal Y}_{b} \,{\cal X}^{(ab)cd,e}{}_f
\, (U^{-1})_e{}^{q} {\cal K}_{[cd]}{}^m  \partial_m    U_q{}^{f}
\nonumber\\
&&{}
-\frac25\,\sqrt{|{\rm g}|}\, {\cal Y}_a {\cal Y}_b \,\eta_{cd} \, M^{ac,bd}  
\;.
\label{MMJJ7}
\eea
in terms of the ${\rm SL}(6)$ twist matrix (\ref{twistfirst}), and the combination
\bea
{\cal X}^{(ab)cd,e}{}_f &=&{\cal X}^{(ab)[cd],e}{}_f~\equiv~
2\,{M}^{je,g(a}  M^{b)h,cd}  {M}_{gh,jf}  -
{M}_{f\alpha}{}^{g(a} M^{b)h,cd}  {M}_{gh}{}^{e\alpha} 
\;,
\label{tensorX}
\eea
of matrix components of (\ref{MD5}). At first view, the structure of this expression 
in no way ressembles the form of (\ref{rhs}), with a far more complicated $y$-dependence
in its first term.
This seemingly jeopardizes the consistency of the reduction of equation (\ref{FF5A}),
which after all should be guaranteed by consistency of the ansatz.
What comes to the rescue is some additional properties of the twist matrix together with
some highly non-trivial non-linear identities among the components of an ${\rm E}_{6(6)}$ matrix.
Namely the last factor in the 
first term of (\ref{MMJJ7}) drastically reduces upon certain index projections 
\bea
(U^{-1})_{a}{}^{q} {\cal K}_{[bc]}{}^m  \partial_m    U_{q}{}^{c}+
(U^{-1})_{b}{}^{q} {\cal K}_{[ac]}{}^m  \partial_m    U_{q}{}^{c} &=&
-\sqrt{2}\,\eta_{ab}
\;,\nonumber\\
(U^{-1})_{a}{}^{q} {\cal K}_{[bc]}{}^m  \partial_m    U_{q}{}^{d}+ 
(U^{-1})_{b}{}^{q} {\cal K}_{[ca]}{}^m  \partial_m    U_{q}{}^{d}+
(U^{-1})_{c}{}^{q} {\cal K}_{[ab]}{}^m  \partial_m    U_{q}{}^{d}
&=& 0
\;,
\label{UJ2}
\eea
as may be verified by explicit computation. Moreover, the tensor ${\cal X}^{(ab)cd,e}{}_f$
defined in (\ref{tensorX}) is of quite restricted nature and satisfies 
\bea
{\cal X}^{(ab)cd,e}{}_f &=& 
{\cal X}^{(ab)[cd,e]}{}_f -\frac25\,\delta_f{}^{[c}\, {\cal X}^{(ab)d]g,e}{}_g
-\frac{2}{45}\,\delta_f{}^{[c} \,{\cal X}^{(ab)d]e,g}{}_g
+\frac19\,\delta_f{}^e\,{\cal X}^{(ab)cd,g}{}_g\;,\qquad
\label{E6identity}
\eea
implying in particular that
\bea
{\cal X}^{(ab)e[c,d]}{}_e
&=&-\frac16\,\,{\cal X}^{(ab)cd,e}{}_e
\;.
\eea
The identity (\ref{E6identity}) is far from obvious and hinges on the 
group properties of the matrix~(\ref{MD5}). It can be verified by choosing an
explicit parametrization of this matrix (e.g.\ as given in \cite{Baguet:2015xha}),
at least with the help of some computer algebra~\cite{Peeters:2006kp,Peeters:2007wn,mathematica}.
Combining this identity with the properties (\ref{UJ2}) of the twist matrix, we conclude that
the first term on the r.h.s.\ of (\ref{MMJJ7}) simplifies according to
\bea
{\cal X}^{(ab)cd,e}{}_f
\, (U^{-1})_e{}^{q} {\cal K}_{[cd]}{}^m  \partial_m    U_q{}^{f}&=&
\frac25\,{\cal X}^{(ab)g(d,e)}{}_g\, (U^{-1})_e{}^{q} {\cal K}_{[fd]}{}^m  \partial_m    U_q{}^{f}
\nonumber\\
&=&
\frac15\,\sqrt{2}\,{\cal X}^{(ab)gd,e}{}_g\, \eta_{de}
\;,
\eea
such that its $y$-dependence reduces to the harmonics ${\cal Y}_a{\cal Y}_b$.

As a consequence, together with (\ref{UJ2}), we conclude
that the penultimate term in (\ref{MMJJ7}) reduces to
\bea
-\frac1{10}\,\sqrt{2}\, \sqrt{|g|}\,{\cal Y}_{a} {\cal Y}_{b} \,{\cal X}^{(ab)cd,e}{}_f\,
(U^{-1})_e{}^{q} {\cal K}_{[cd]}{}^l  \partial_l    U_q{}^{f}
&=&
-\frac1{25}\,\sqrt{|g|}\, {\cal Y}_{a} {\cal Y}_{b} \,{\cal X}^{(ab)gc,d}{}_g\,
\eta_{cd}
\;.\quad
\eea
Together with (\ref{rhs_exp}), equation (\ref{FF5A}) then eventually reduces to 
\bea
D_{[\mu} \Lambda_{\nu\rho\sigma\tau]}
&=& 
-\frac1{80}\, {\cal Y}_{a}  {\cal Y}^{b}
\,\sqrt{|{\rm g}|}\,\varepsilon_{\mu\nu\rho\sigma\tau} D_{\lambda}
\left({M}^{N\,ac} \,{D}^\lambda {M}_{bc,N} \right) 
\nonumber\\
&&{} 
+\frac{1}{40}\,
{\cal Y}_a {\cal Y}^b \, 
\sqrt{|{\rm g}|}\,\varepsilon_{\mu\nu\rho\sigma\tau}\,{F}^{\kappa\lambda\,N}\left(
{M}_{bc,N}  {F}_{\kappa\lambda}{}^{ac}
-\frac12\,\sqrt{10}\,\varepsilon_{\alpha\beta}\,\eta_{db}\,M^{d\,\alpha}{}_N\,
{B}_{\kappa\lambda}{}^{a\,\beta}  
\right)
\nonumber\\
&&{}
+\frac1{100}\,\sqrt{|{\rm g}|}\,\varepsilon_{\mu\nu\rho\sigma\tau}\,
{\cal Y}_{a} {\cal Y}^{b} \left(
10\,M^{ac,fd}  
+ {\cal X}^{(af)ec,d}{}_e
\right) \eta_{cd} \eta_{bf}
\nonumber\\
&&{}
+\frac1{32}\,\sqrt{2}\, 
 \,\varepsilon_{abcdef}\,  {F}_{[\mu\nu}{}^{ab} F_{\rho\sigma}{}^{cd}  A_{\tau]}{}^{ef}
+ \frac1{16}\,
 {F}_{[\mu\nu}{}^{ab} A_\rho{}^{cd} A_{\sigma}{}^{ef}  A_{\tau]}{}^{gh} \,\varepsilon_{abcdeh} \eta_{fh}
 \nonumber\\
&&{}
+\frac1{40}\,\sqrt{2}\,\, A_{[\mu}{}^{ab} A_{\nu}{}^{cd}A_\rho{}^{ef} A_\sigma{}^{gh} A_{\tau]}{}^{ij}\,
 \varepsilon_{abcegi}\,\eta_{df}\eta_{hj}
\;,
\label{FF5F}
\eea
such that the $y$-dependence of the entire equation 
organizes into the form (\ref{rhs}).
Now the $x$-dependent coefficient of the traceless combination
$\left({\cal Y}_a{\cal Y}_b - \frac16\,\eta_{ab}\right)$ precisely reproduces
the $D=5$ scalar equations of motion (\ref{D5scalarEOM}).
In particular, the third line of (\ref{FF5F}) coincides with the ${\rm SL}(6)$ variation
of the scalar potential (\ref{potential}).
This match requires additional non-trivial relations
among the components of an ${\rm E}_{6(6)}$ matrix (\ref{MD5})
\bea
\eta_{ef}    {M}_{d\alpha}{}^{h(a}  {M}^{b)c,de}   {M}^{f\alpha}{}_{ch}   &=&
\eta_{ef}      {M}_{g\alpha}{}^{de} {M}^{fc,g(a}  M^{b)\alpha}{}_{cd}  
\;,
\\
\eta_{ef} {M}^{de,c(a} {M}^{b)\gamma,f\alpha}   {M}_{d\alpha,c\gamma}&=&
2\, \eta_{ef} \,{M}^{de,c(a}  {M}^{b)h,fg} {M}_{dg,ch} 
+\eta_{ef}  {M}_{d\alpha}{}^{h(a}  {M}^{b)c,de}   {M}^{f\alpha}{}_{ch} 
\;,
\nonumber
\eea
which can be proven similar to (\ref{E6identity}).
From these it is straightforward to deduce that
\bea
{\cal X}^{(af)ec,d}{}_e &=& -\frac43\, {M}^{de,c(a}  {M}^{b)h,fg} {M}_{dg,ch}  \eta_{ef}  
-\frac13\,  \eta_{ef} {M}^{de,c(a} {M}^{b)\gamma,f\alpha}   {M}_{d\alpha,c\gamma}     
\nonumber\\
&&{}
+ \frac23 \,  \eta_{de} {M}^{cd,g(a}  M^{b\alpha}{}_{cf}  {M}_{g\alpha}{}^{ef} 
+ \frac23\,\eta_{ef}  {M}^{de,c(a}     {M}^{b)h}{}_{d\alpha} {M}^{f\alpha}{}_{ch}  
\;,
\eea
thus matching the expression obtained from variation of the scalar potential in (\ref{D5scalarEOM}).
As a consequence, the $y$-dependent part of 
equation (\ref{FF5F}) vanishes on-shell, such that the
equation reduces to
\bea
D_{[\mu} \Lambda_{\nu\rho\sigma\tau]}
&=& 
-\frac1{480}\, 
\,\sqrt{|{\rm g}|}\,\varepsilon_{\mu\nu\rho\sigma\tau} D_{\lambda}
\left({M}^{N\,ac} \,{D}^\lambda {M}_{ac,N} \right) 
\nonumber\\
&&{} 
+\frac{1}{240}\, 
\sqrt{|{\rm g}|}\,\varepsilon_{\mu\nu\rho\sigma\tau}\,{F}^{\kappa\lambda\,N}\left(
{M}_{ab,N}  {F}_{\kappa\lambda}{}^{ab}
-\frac12\,\sqrt{10}\,\varepsilon_{\alpha\beta}\,\eta_{ab}\,M^{a\,\alpha}{}_N\,
{B}_{\kappa\lambda}{}^{b\,\beta}  
\right)
\nonumber\\
&&{}
+\frac1{600}\,\sqrt{|{\rm g}|}\,\varepsilon_{\mu\nu\rho\sigma\tau}
 \left(
10\,M^{ac,fd}  
+ {\cal X}^{(af)ec,d}{}_e
\right) \eta_{cd} \eta_{af}
\nonumber\\
&&{}
+\frac1{32}\,\sqrt{2}\, 
 \,\varepsilon_{abcdef}\,  {F}_{[\mu\nu}{}^{ab} F_{\rho\sigma}{}^{cd}  A_{\tau]}{}^{ef}
+ \frac1{16}\,
 {F}_{[\mu\nu}{}^{ab} A_\rho{}^{cd} A_{\sigma}{}^{ef}  A_{\tau]}{}^{gh} \,\varepsilon_{abcdeh} \eta_{fh}
 \nonumber\\
&&{}
+ \frac1{40}\,\sqrt{2}\,\, A_{[\mu}{}^{ab} A_{\nu}{}^{cd}A_\rho{}^{ef} A_\sigma{}^{gh} A_{\tau]}{}^{ij}\,
 \varepsilon_{abcegi}\,\eta_{df}\eta_{hj}
\;.
\label{LambdaA}
\eea
This equation can be integrated to yield the function $\Lambda_{\mu\nu\rho\rho}$.
This yields the last missing part in the reduction ansatz of the IIB four form (\ref{C4})
and establishes the full type IIB self-duality equation.

\end{appendix}


\begin{thebibliography}{10}

\bibitem{Duff:1984hn}
M.~Duff, B.~Nilsson, C.~Pope, and N.~Warner, { On the consistency of the
  {K}aluza-{K}lein ansatz},  { Phys.Lett.} { B149} (1984)
90.

\bibitem{Cvetic:2000dm}
M.~Cvetic, H.~Lu, and C.~N. Pope, { Consistent {K}aluza-{K}lein sphere
  reductions},  { Phys. Rev.} { D62} (2000) 064028,
[\href{http://xxx.lanl.gov/abs/hep-th/0003286}{{\tt hep-th/0003286}}].

\bibitem{Gunaydin:1984qu}
  M.~Gunaydin, L.~J.~Romans and N.~P.~Warner,
  Gauged $N=8$ supergravity in five-dimensions,
  Phys.\ Lett.\ B {154} (1985) 268.

\bibitem{Pernici:1985ju}
  M.~Pernici, K.~Pilch and P.~van Nieuwenhuizen,
  Gauged $N=8$, $D=5$ supergravity,
  Nucl.\ Phys.\ B {259} (1985) 460.

\bibitem{Gunaydin:1985cu}
M.~G{\"u}naydin, L.~J. Romans, and N.~P. Warner, { Compact and noncompact
  gauged supergravity theories in five-dimensions},  { Nucl. Phys.} { B272}
  (1986)
598.

\bibitem{deWit:1986iy}
B.~de~Wit and H.~Nicolai, { The consistency of the {$S^7$} truncation in
  {$D=11$} supergravity},  { Nucl.Phys.} { B281} (1987)
211.

\bibitem{Nastase:1999kf}
H.~Nastase, D.~Vaman, and P.~van Nieuwenhuizen, { Consistency of the {AdS}$_7
  \times {S}^4$ reduction and the origin of self-duality in odd dimensions},  {
  Nucl. Phys.} { B581} (2000) 179--239,
[\href{http://xxx.lanl.gov/abs/hep-th/9911238}{{\tt hep-th/9911238}}].

\bibitem{Cvetic:1999xp}
M.~Cvetic, M.~Duff, P.~Hoxha, J.~T. Liu, H.~Lu, J.~Lu, R.~Martinez-Acosta,
  C.~Pope, H.~Sati, and T.~A. Tran, { Embedding {AdS} black holes in
  ten-dimensions and eleven-dimensions},  { Nucl.Phys.} { B558} (1999) 96--126,
[\href{http://xxx.lanl.gov/abs/hep-th/9903214}{{\tt hep-th/9903214}}].

\bibitem{Lu:1999bw}
H.~Lu, C.~Pope, and T.~A. Tran, { Five-dimensional {$N=4$, $SU(2) \times U(1)$}
  gauged supergravity from type {IIB}},  { Phys.Lett.} { B475} (2000) 261--268,
[\href{http://xxx.lanl.gov/abs/hep-th/9909203}{{\tt hep-th/9909203}}].

\bibitem{Cvetic:2000nc}
M.~Cvetic, H.~Lu, C.~Pope, A.~Sadrzadeh, and T.~A. Tran, { Consistent ${SO}(6)$
  reduction of type {IIB} supergravity on {$S^5$}},  { Nucl.Phys.} { B586}
  (2000) 275--286,
[\href{http://xxx.lanl.gov/abs/hep-th/0003103}{{\tt hep-th/0003103}}].

\bibitem{Khavaev:1998fb}
  A.~Khavaev, K.~Pilch and N.~P.~Warner,
  New vacua of gauged N=8 supergravity in five-dimensions,
  Phys.\ Lett.\ B {487} (2000) 14
  [hep-th/9812035].

\bibitem{Nastase:2000tu}
  H.~Nastase and D.~Vaman,
  On the nonlinear KK reductions on spheres of supergravity theories,
  Nucl.\ Phys.\ B {583} (2000) 211
  [hep-th/0002028].

\bibitem{Pilch:2000ue}
K.~Pilch and N.~P. Warner, { {$N=2$} supersymmetric {RG} flows and the {IIB}
  dilaton},  { Nucl.Phys.} { B594} (2001) 209--228,
[\href{http://xxx.lanl.gov/abs/hep-th/0004063}{{\tt hep-th/0004063}}].

\bibitem{Cassani:2010uw} 
  D.~Cassani, G.~Dall'Agata and A.~F.~Faedo,
  Type IIB supergravity on squashed Sasaki-Einstein manifolds,
  JHEP { 1005} (2010) 094,
  [{\tt 1003.4283}].

\bibitem{Liu:2010sa} 
  J.~T.~Liu, P.~Szepietowski and Z.~Zhao,
  Consistent massive truncations of IIB supergravity on Sasaki-Einstein manifolds,
  Phys. Rev. D {81} (2010) 124028, 
 [{\tt 1003.5374}].

\bibitem{Gauntlett:2010vu} 
  J.~P.~Gauntlett and O.~Varela,
  Universal Kaluza-Klein reductions of type IIB to N=4 supergravity in five dimensions,
  JHEP {1006} (2010) 081,
   [{\tt 1003.5642}].
  
\bibitem{Skenderis:2010vz}
K.~Skenderis, M.~Taylor, and D.~Tsimpis, { A consistent truncation of {IIB}
  supergravity on manifolds admitting a {S}asaki-{E}instein structure},  {
  JHEP} { 1006} (2010) 025,
[\href{http://xxx.lanl.gov/abs/1003.5657}{{\tt 1003.5657}}].

\bibitem{Guarino:2015jca} 
  A.~Guarino, D.~L.~Jafferis and O.~Varela,
  { The String Origin of Dyonic ${\mathcal N} = 8$ Supergravity and its Simple Chern-Simons Duals}, 
  [\href{http://xxx.lanl.gov/abs/1504.08009}{{\tt 1504.08009}}].

\bibitem{Hohm:2014qga}
O.~Hohm and H.~Samtleben, { Consistent {K}aluza-{K}lein truncations via
  exceptional field theory},  { JHEP} { 1501} (2015) 131,
[\href{http://xxx.lanl.gov/abs/1410.8145}{{\tt 1410.8145}}].

\bibitem{Hull:1988jw}
C.~M. Hull and N.~P. Warner, { Noncompact gaugings from higher dimensions},  {
  Class. Quant. Grav.} { 5} (1988)
1517.

\bibitem{Cvetic:2004km}
M.~Cvetic, G.~Gibbons, and C.~Pope, { Ghost free de {S}itter supergravities as
  consistent reductions of string and {M} theory},  { Nucl.Phys.} { B708}
  (2005) 381--410,
[\href{http://xxx.lanl.gov/abs/hep-th/0401151}{{\tt hep-th/0401151}}].

\bibitem{Baron:2014bya}
W.~H. Baron and G.~Dall'Agata, { {Uplifting non-compact gauged
  supergravities}},  { JHEP} { 1502} (2015) 003,
[\href{http://xxx.lanl.gov/abs/1410.8823}{{\tt 1410.8823}}].

\bibitem{Hohm:2013pua}
O.~Hohm and H.~Samtleben, { Exceptional form of ${D}=11$ supergravity},  {
  Phys.Rev.Lett.} { 111} (2013) 231601,
[\href{http://xxx.lanl.gov/abs/1308.1673}{{\tt 1308.1673}}].

\bibitem{Hohm:2013vpa}
O.~Hohm and H.~Samtleben, { Exceptional field theory {I}: ${E}_{6(6)}$
  covariant form of {M}-theory and type {IIB}},  { Phys.Rev.} { D89} (2014)
  066016,
[\href{http://xxx.lanl.gov/abs/1312.0614}{{\tt 1312.0614}}].

\bibitem{Hohm:2013uia}
O.~Hohm and H.~Samtleben, { Exceptional field theory {II}: {E}$_{7(7)}$},  {
  Phys.Rev.} { D89} (2014) 066017,
[\href{http://xxx.lanl.gov/abs/1312.4542}{{\tt 1312.4542}}].

\bibitem{Hohm:2014fxa}
O.~Hohm and H.~Samtleben, { Exceptional field theory {III}: ${E}_{8(8)}$},  {
  Phys.Rev.} { D90} (2014) 066002,
[\href{http://xxx.lanl.gov/abs/1406.3348}{{\tt 1406.3348}}].

\bibitem{Berman:2010is}
D.~S. Berman and M.~J. Perry, { Generalized geometry and {M} theory},  { JHEP}
  { 1106} (2011) 074,
[\href{http://xxx.lanl.gov/abs/1008.1763}{{\tt 1008.1763}}].

\bibitem{Coimbra:2011ky}
A.~Coimbra, C.~Strickland-Constable, and D.~Waldram, { {$E_{d(d)} \times
  \mathbb{R}^+$ generalised geometry, connections and M theory}},  { JHEP} {
  1402} (2014) 054,
[\href{http://xxx.lanl.gov/abs/1112.3989}{{\tt 1112.3989}}].

\bibitem{Aldazabal:2013mya}
G.~Aldazabal, M.~Gra{\~n}a, D.~Marqu{\'e}s, and J.~Rosabal, { {Extended
  geometry and gauged maximal supergravity}},  { JHEP} { 1306} (2013) 046,
[\href{http://xxx.lanl.gov/abs/1302.5419}{{\tt 1302.5419}}].

\bibitem{Hohm:2015xna}
O.~Hohm and Y.-N. Wang, { Tensor hierarchy and generalized {C}artan calculus in
  {SL}(3) $\times$ {SL}(2) exceptional field theory},  { JHEP} { 1504} (2015)
  050,
[\href{http://xxx.lanl.gov/abs/1501.01600}{{\tt 1501.01600}}].

\bibitem{Siegel:1993th}
W.~Siegel, { {Superspace duality in low-energy superstrings}},  { Phys.Rev.} {
  D48} (1993) 2826--2837,
[\href{http://xxx.lanl.gov/abs/hep-th/9305073}{{\tt hep-th/9305073}}].

\bibitem{Hull:2009mi}
C.~Hull and B.~Zwiebach, { Double field theory},  { JHEP} { 0909} (2009) 099,
[\href{http://xxx.lanl.gov/abs/0904.4664}{{\tt 0904.4664}}].

\bibitem{Hohm:2010jy}
O.~Hohm, C.~Hull, and B.~Zwiebach, { {Background independent action for double
  field theory}},  { JHEP} { 1007} (2010) 016,
[\href{http://xxx.lanl.gov/abs/1003.5027}{{\tt 1003.5027}}].

\bibitem{Hohm:2010pp}
O.~Hohm, C.~Hull, and B.~Zwiebach, { {Generalized metric formulation of double
  field theory}},  { JHEP} { 1008} (2010) 008,
[\href{http://xxx.lanl.gov/abs/1006.4823}{{\tt 1006.4823}}].

\bibitem{Cvetic:2003jy}
M.~Cvetic, G.~Gibbons, H.~Lu, and C.~Pope, { {Consistent group and coset
  reductions of the bosonic string}},  { Class.Quant.Grav.} { 20} (2003)
  5161--5194,
[\href{http://xxx.lanl.gov/abs/hep-th/0306043}{{\tt hep-th/0306043}}].

\bibitem{Scherk:1979zr}
J.~Scherk and J.~H. Schwarz, { How to get masses from extra dimensions},  {
  Nucl. Phys.} { B153} (1979)
61--88.

\bibitem{Berman:2012vc}
D.~S. Berman, M.~Cederwall, A.~Kleinschmidt, and D.~C. Thompson, { {The gauge
  structure of generalised diffeomorphisms}},  { JHEP} { 1301} (2013) 064,
[\href{http://xxx.lanl.gov/abs/1208.5884}{{\tt 1208.5884}}].

\bibitem{Baguet:2015xha}
A.~Baguet, O.~Hohm, and H.~Samtleben, { {E$_{6(6)}$} exceptional field theory:
  Review and embedding of type {IIB}}, 
  Contribution to the Proceedings of the Workshop on Quantum Fields and Strings, Corfu 2014,
  PoS (CORFU2014) 133, 
  [\href{http://xxx.lanl.gov/abs/1506.01065}{{\tt 1506.01065}}].

\bibitem{Lee:2014mla}
K.~Lee, C.~Strickland-Constable, and D.~Waldram, { {Spheres, generalised
  parallelisability and consistent truncations}},
\href{http://xxx.lanl.gov/abs/1401.3360}{{\tt 1401.3360}}.

\bibitem{Ciceri:2014wya}
F.~Ciceri, B.~de~Wit, and O.~Varela, { {IIB} supergravity and the {E}$_{6(6)}$
  covariant vector-tensor hierarchy},  { JHEP} { 1504} (2015) 094,
[\href{http://xxx.lanl.gov/abs/1412.8297}{{\tt 1412.8297}}].

\bibitem{deWit:2004nw}
B.~de~Wit, H.~Samtleben, and M.~Trigiante, { The maximal ${D} = 5$
  supergravities},  { Nucl. Phys.} { B716} (2005) 215--247,
[\href{http://xxx.lanl.gov/abs/hep-th/0412173}{{\tt hep-th/0412173}}].

\bibitem{Townsend:1983xs}
P.~K. Townsend, K.~Pilch, and P.~van Nieuwenhuizen, { Selfduality in odd
  dimensions},  { Phys. Lett.} { 136B} (1984)
38.

\bibitem{Schwarz:1983wa}
J.~H. Schwarz and P.~C. West, { {Symmetries and transformations of chiral
  ${N}=2$ ${D}=10$ supergravity}},  { Phys. Lett.} { B126} (1983)
301.

\bibitem{Schwarz:1983qr}
J.~H. Schwarz, { Covariant field equations of chiral {$N=2$, $D=10$}
  supergravity},  { Nucl.Phys.} { B226} (1983)
269.

\bibitem{Howe:1983sra}
P.~S. Howe and P.~C. West, { {The complete ${N}=2$, ${D}=10$ supergravity}},  {
  Nucl. Phys.} { B238} (1984)
181.

\bibitem{Godazgar:2013pfa}
H.~Godazgar, M.~Godazgar, and H.~Nicolai, { {Nonlinear Kaluza-Klein theory for
  dual fields}},  { Phys.Rev.} { D88} (2013) 125002,
[\href{http://xxx.lanl.gov/abs/1309.0266}{{\tt 1309.0266}}].

\bibitem{deWit:2008ta}
B.~de~Wit, H.~Nicolai, and H.~Samtleben, { Gauged supergravities, tensor
  hierarchies, and {M}-theory},  { JHEP} { 0802} (2008) 044,
[\href{http://xxx.lanl.gov/abs/arXiv:0801.1294}{{\tt arXiv:0801.1294}}].

\bibitem{Duff:1986hr}
M.~Duff, B.~Nilsson, and C.~Pope, { Kaluza-{K}lein supergravity},  {
  Phys.Rept.} { 130} (1986)
1--142.

\bibitem{Musaev:2014lna}
E.~Musaev and H.~Samtleben, { Fermions and supersymmetry in {E}$_{6(6)}$
  exceptional field theory},  { JHEP} { 1503} (2015) 027,
[\href{http://xxx.lanl.gov/abs/1412.7286}{{\tt 1412.7286}}].

\bibitem{Peeters:2006kp}
K.~Peeters, { {A field-theory motivated approach to symbolic computer
  algebra}},  { Comput. Phys. Commun.} { 176} (2007) 550--558,
[\href{http://xxx.lanl.gov/abs/cs/0608005}{{\tt cs/0608005}}].

\bibitem{Peeters:2007wn}
K.~Peeters, { Introducing {C}adabra: {A} symbolic computer algebra system for
  field theory problems},
\href{http://xxx.lanl.gov/abs/hep-th/0701238}{{\tt hep-th/0701238}}.

\bibitem{mathematica}
{Wolfram Research, Inc.}, { Mathematica}.
\newblock Champaign, Illinois, version 8.0~ed., 2010.

\end{thebibliography}

\providecommand{\href}[2]{#2}\begingroup\raggedright\endgroup

\end{document}